\documentclass[aps,pre,preprint,nofootinbib,showpacs,eqsecnum,amsmath,
titlepage,tightenlines]{revtex4}

\usepackage{graphicx}

\begin{document}
\preprint{OKHEP--02--10}

\title{Does the Transverse Electric Zero Mode Contribute to the Casimir Effect
 for a Metal?}

\author{J. S. H{\o}ye}
\email{johan.hoye@phys.ntnu.no}
\affiliation{Department of Physics, Norwegian University of Science and
Technology, N-7491, Trondheim, Norway}

\author{I. Brevik}
\email{iver.h.brevik@mtf.ntnu.no} 
\author{J. B.  Aarseth}
\email{jan.b.aarseth@mtf.ntnu.no}
\affiliation{Department of Energy and Process
Engineering, Norwegian University of Science and
Technology, N-7491, Trondheim, Norway}

\author{K. A. Milton}
\email{milton@nhn.ou.edu}
\affiliation{Department of Physics and Astronomy, The University of Oklahoma,
Norman, OK 73019 USA}

\date{\today}

\begin{abstract}
The finite temperature Casimir free energy, entropy, and internal energy
are considered anew for a conventional parallel-plate configuration, in
the light of current discussions in the literature. In the case of an
``ideal" metal, characterized by a refractive index equal to infinity
for all frequencies, we recover, via a somewhat unconventional method,
conventional results for the temperature dependence, meaning that the
zero-frequency transverse electric mode contributes the same 
as the transverse magnetic mode. For a real metal, however, approximately
obeying the Drude dispersive model at low frequencies, we find that the
zero-frequency transverse electric mode does not contribute at all.
This would appear to lead to an observable temperature dependence
and a violation of the third law of thermodynamics.
It had been suggested that the source of the difficulty was
 the behaviour of the reflection coefficient for
perpendicular polarization but we show that this
is not the case. By introducing a simplified
model for the Casimir interaction, consisting of two harmonic
oscillators interacting via a third one, 
we illustrate the behavior of the transverse
electric field. 
 Numerical results are presented based on the refractive index for gold.
A linear temperature correction to the Casimir force between parallel
plates is indeed found which should be observable in room-temperature 
experiments, but this does not entail any thermodynamic inconsistency.

\end{abstract}


\maketitle

\section{Introduction}
\label{I}
In spite of the numerous treatises on the Casimir effect during the
past decade---for some books and review papers see, for instance,
Milton \cite{milton01}, Mostepanenko and Trunov \cite{mostepanenko97},
Milonni~\cite{milonni94}, Plunien et al.~\cite{plunien86}, Bordag et
al.~\cite{bordag01}---it is somewhat surprising that such a basic issue
as the temperature dependence of this effect is still unclear and has recently
given rise to a lively discussion. This issue is not
restricted to the case of curvilinear geometry, but is present even in
the simplest conventional geometry of two parallel metal plates
separated by a gap of width $a$. Thus Klimchitskaya and
Mostepanenko in their detailed investigation \cite{klimchitskaya01},
and also Bordag et al.~\cite{bordag00}, and Fischbach et al.~\cite{fischbach01},
have argued that the Drude dispersion relation for
a frequency-dispersive medium leads to inconsistencies in the sense
that the reflection coefficient $r_2$ for perpendicular
polarization (the TE mode)
becomes discontinuous as the imaginary frequency $\zeta=
-i\omega$ goes to zero. As is well known, the Drude dispersion relation reads
for imaginary frequencies
\begin{equation}
\varepsilon(i\zeta)=1+\frac{\omega_p^2}{\zeta (\zeta+\nu)},
\label{1}
\end{equation}
where $\omega_p$ is the plasma frequency and $\nu$ the relaxation
frequency. (Usually, $\nu$ is taken to be a constant, equal to its 
room-temperature value.) 
The mentioned authors, instead of the Drude relation, give
preference to the plasma dispersion relation, since no such
discontinuity is then encountered. (In Ref.~\cite{bezerra02}, the plasma 
relation together with the so-called surface impedance approach is argued to 
be the method best suited to describe the thermal Casimir force between real 
metals.)
The plasma relation is
\begin{equation}
\varepsilon(i\zeta)=1+\frac{\omega_p^2}{\zeta^2}.
\label{2}
\end{equation}
The arguments in Refs.~\cite{klimchitskaya01,bordag00,fischbach01,bezerra02} 
are interesting, since they raise doubts not only about the applicability
of the Drude model as such, but even more, doubt about the
applicability of the fundamental Lifshitz formula at low temperatures
(see, for instance, Ref.~\cite{lifshitz80}).

The essence of the problem appears to be the following: 
For a metal, does the transverse
electric (TE) mode contribute to the Casimir effect in the limit of
zero frequency, corresponding to Matsubara integer $m=0$? It is
precisely for this mode that the purported discontinuity  of the
reflection coefficient $r_2$, mentioned above, can occur. The problem
is most acute in the high $T$ regime (the $m=0$ contribution becomes
increasingly important as $T$ increases), but is present at moderate
and low temperatures as well. The conventional recipe for handling the
two-limit problem for a metal, $n=\sqrt{\varepsilon} \rightarrow
\infty$, $m\rightarrow 0$, has been to take the limits in the following
order:
\begin{enumerate}
\item\label{i}  Set first $\varepsilon =\infty$;
\item\label{ii}  then take the limit $m=0$.
\end{enumerate}
This way of proceeding was advocated in the early paper of Schwinger,
$\rm{DeRaad}$, and Milton \cite{schwinger78} (we will call it the
SDM prescription), and was followed also in one of the
recent papers by some of the current authors \cite{hoye01}, 
and in Milton's recent book 
\cite{milton01}.  It seems to escaped recent notice that the physical
basis for this prescription, namely the necessity of enforcing the correct
electrostatic boundary conditions, was explicitly stated in 
Ref.~\cite{schwinger78}.

Bostr{\"o}m and Sernelius \cite{bostrom00} seem to have been the first
to inquire whether this prescription is right: They argued that in view
of a realistic dispersion relation at low frequencies the $m=0$ TE
mode should \textit{not\/} contribute. And three of the present
authors arrived  recently at
the same conclusion, in two papers dealing with the case of two
concentric spherical surfaces \cite{brevik02,brevik02a}.

The Bostr{\"o}m-Sernelius paper gave rise to a heated debate in the
literature \cite{bordag00,lamoreaux00,sernelius01,bordag01a} on the
role of the $m=0$ TE mode for a metal. The advent of accurate
experiments in recent years, by Lamoreaux \cite{lamoreaux97}, Mohideen
et al.~\cite{mohideen98,roy99,harris00,chen02}, Ederth \cite{ederth00},
Chan et al.~\cite{chan01}, and Bressi et al.~\cite{bressi02} (cf.~also
the recent review paper of Lambrecht and Reynaud \cite{lambrecht02}),
represents important progress in this field. Especially the
experiment of Bressi et al.\ is of interest in the present context,
since it deals directly with the Casimir force between metal surfaces
that are parallel, and so avoids use of complicating factors such as
the proximity force theorem \cite{blocki77}, which nevertheless
seems well understood. This experiment is fraught
with experimental difficulties (related to keeping the plates
sufficiently parallel), so the accuracy is claimed by the authors to be moderate (15\%), but
it is to be hoped that this accuracy will soon be improved. Several
other related papers have appeared recently, discussing the
interpretation of the mentioned experiments as well as more general
aspects of finite temperature Casimir theory
\cite{lamoreaux98,lambrecht00,genet02,svetovoy00,barton01,feinberg01,bezerra02a}.

Our purpose in the present paper is to analyze the Casimir temperature
problem anew, assuming conventional parallel-plate geometry from the
outset, therewith avoiding the spherical Bessel functions that become
necessary if spherical geometry is contemplated. In particular, we will
focus attention on the $m=0$ TE mode. Let us summarize our results:

It is useful to distinguish between two different classes of metals.
The first class, which we will call ``ideal" metals, is characterized by
a refractive index $n=\sqrt{\varepsilon}=\infty$ for all
frequencies. It implies that the reflection coefficient $r_2$ mentioned
above is unity for all $\zeta$. This corresponds to the traditional recipe
\ref{i} and \ref{ii} above when handling the two-limit problem for metals. It
means that the $m=0$ TE mode contributes to the Casimir force
just the same amount as does the transverse magnetic (TM) mode.

The obvious drawback of this ``ideal" metal is that it does not occur in
nature. And this brings us to the second class, which is the one of
real metals, in which case we must observe an appropriate dispersion
relation, especially at low frequencies.  It is most commonly assumed
that the most appropriate dispersion when $\zeta \rightarrow 0$ is the
Drude relation, Eq.~(\ref{1}). As we will show, the Drude model implies that
the $m=0$ TE mode does \textit{not} contribute. The total $m=0$ free
energy for a real metal becomes accordingly one half of the
conventional expression. In contradistinction to recent statements in
the literature \cite{klimchitskaya01,bordag00,fischbach01} we find that
there exists no physical difficulty or ambiguity associated with the vanishing 
coefficient $r_2$ at $\zeta=0$. This is so because $r_2$ goes to zero smoothly 
when $\zeta \rightarrow 0$, as long as the transverse wave vector 
$\mathbf{  k}_\perp$ is nonvanishing. (If $\mathbf{k}_\perp$ is precisely zero, 
there occurs a singularity in the reflection coefficient, but this has no 
physical importance since this point is of measure zero in the integral over 
$\mathbf{k}_\perp$.)
 Our present results are in agreement with Refs.~\cite{brevik02,brevik02a}, 
as well as with Bostr{\"o}m and Sernelius \cite{bostrom00}. 

A different view has recently been put forward by Torgerson and Lamoreaux 
\cite{torgerson}. They argue that 
the Drude-model behavior does not accurately represent the
TE zero mode, which necessarily has a vanishing tangential component
at the surface of a perfect conductor.  They point to the necessity of taking 
the finite thickness of the metallic coatings into account. Their arguments 
seem to imply that the conventional temperature dependence is correct.
 However, in our opinion electrostatic 
 considerations of this kind do not solve the zero temperature problem;
 what is required to incorporate temperature dependence is an analytic
 continuation into imaginary frequencies of Green's functions referring
 to nonzero wavenumber.

Before embarking on the calculations let us emphasize the following
point: The occurrence of the $m=0$ mode only once instead of twice
is understandable physically. This mode is precisely the TM static
mode, corresponding to the electric field being perpendicular 
to the two metal plates. It is the natural ground-state mode
present when $\zeta =0$. Actually, in Sec.~III of Ref.~\cite{hoye01} we
showed how the uniqueness of the static mode emerges naturally, using 
statistical mechanical considerations.

The outline of our paper is the following.
 In the next section we show why the exclusion of the TE zero mode
 seems to lead to an observable temperature correction to the force between
 real metal plates, and worse, seems to imply a violation of the
 third law of thermodynamics.
    In Sec.~\ref{IV} we expand on the situation of
an ``ideal" metal in the sense described above, and calculate the
Casimir free energy, entropy, and internal energy via a
somewhat unconventional
route. Equivalence with earlier results is demonstrated. In Sec.~\ref{V} we
introduce a new and simplified model to illustrate the Casimir problem, based
essentially on statistical mechanics. In this model the system is
replaced by two harmonic oscillators (the two media) that interact via
a third oscillator (the electromagnetic field). 
Depending upon the form of the interaction we then have two situations.
The first is the one where the induced interaction (or free energy),
which is negative, increases linearly in magnitude with temperature
in the classical limit. The other situation, which is more unexpected,
is where the induced interaction vanishes in the classical limit.
These two situations can be regarded as analogous to the behavior of the 
TM and TE modes. We also consider
a strongly simplified case of real metals, and show how in such a case
the contribution to entropy goes to zero smoothly as $T\rightarrow 0$. 
Arguing on basis of the Euler-Maclaurin formula we find this to be a 
general property (except in the idealized metal limit).
 We then go on to present numerical results based on the dispersion
relation for gold, and obtain results qualitatively in accord with our
analytical model. 
In the Appendices the smoothness of the
reflection coefficient $r_2$, and of the TE Green's function,
in the limit $\zeta \rightarrow 0$ is explicitly demonstrated.
We also discuss the temperature dependence of the relaxation frequency,
$\nu(T)$.
We conclude that a linear temperature dependence should be
observable in room temperature experiments. 

In this paper we use natural units, $\hbar=c=k_B=1$.

\section{Temperature Effect for Metal Plates}
\label{II}
We begin by reviewing how temperature effects are incorporated into the
expression for the force between parallel dielectric (or conducting)
plates separated by a distance $a$.
To obtain the finite temperature Casimir force from the zero-temperature
expression, one conventionally makes the following substitution in the
imaginary frequency,
\begin{subequations}
\label{AB}
\begin{equation}
\zeta\to\zeta_m=\frac{2\pi m}{\beta},\label{A}
\end{equation}
and replaces the integral over frequencies by a sum,
\begin{equation}
\int_{-\infty}^\infty \frac{d\zeta}{2\pi}\to\frac1\beta\sum_{m=-\infty}^\infty.
\label{B}
\end{equation}
\end{subequations}
This reflects the requirement that thermal Green's functions be periodic
in imaginary time with period $\beta$ \cite{ms}.
  Suppose we write the finite-temperature force/area as
  [for the explicit form, see Eq.~(\ref{3}) below]
\begin{equation}
{\mathcal   F}^T=\sum_{m=0}^\infty {}'f_m,
\label{primedsum}
\end{equation}
where the prime on the summation sign means that the $m=0$ term is counted
with half weight.
To get the low temperature limit, one can  use the Euler-Maclaurin
(EM) sum  formula,
\begin{equation}
\sum_{k=0}^\infty f(k)=\int_0^\infty f(k)\,dk+\frac{1}{2}f(0)-\sum_{q=
1}^\infty \frac{B_{2q}}{(2q)!} f^{(2q-1)}(0),
\label{22}
\end{equation}
where $B_n$ is the $n$th Bernoulli number.
This means here, with half-weight for the $m=0$ term,
\begin{equation}
{\mathcal   F}^T=\int_0^\infty f(m)\, dm-\frac12 f(0)+\frac12f(0)-\sum_{k=1}^\infty
\frac{B_{2k}}{(2k)!}f^{(2k-1)}(0).
\label{em}
\end{equation}
It is noteworthy that the terms involving $f(0)$ cancel in Eq.~(\ref{em}).
The reason for this is that the EM formula equates an integral
to its trapezoidal-rule approximation plus a series of corrections;
thus the $1/2$ for $m=0$ in Eq.~(\ref{primedsum}) is built in automatically.
For a perfect conductor
\begin{equation}
f(x)=-\frac2{\pi\beta}\int_{2\pi x/\beta}^\infty q^2\,dq\frac1{e^{2qa}-1}.
\end{equation}
Of course, the integral in Eq.~(\ref{em})
 is just the inverse of the finite-temperature
prescription (\ref{B}),
and gives the zero-temperature result.   The only nonzero odd
derivative occurring is 
\begin{equation}
f'''(0)=-\frac{16\pi^2}{\beta^4},
\label{stefan}
\end{equation}
which gives a Stefan's law type of term, seen in Eq.~(\ref{linterm}) below.

The problem is that the EM formula only applies if $f(m)$ is continuous.
If we follow the  argument of Ref.~\cite{bostrom00,brevik02,brevik02a},
 and take the $\epsilon_{1,2}\to\infty$ limit
at the end ($\epsilon_{1,2}$ are the permittivities of the
two parallel dielectric slabs), this is not the case, and for the TE mode
\begin{subequations}
\begin{eqnarray}
f_0&=&0,\\
f_m&=&-\frac{\zeta(3)}{4\pi\beta a^3},\quad 0<\frac{2\pi a m}{\beta}\ll1.
\end{eqnarray}
\end{subequations}
Then we have to modify the argument as follows:
\begin{eqnarray}
{\mathcal   F}^T&=&\sum_{m=0}^\infty{}'f_m=\sum_{m=1}^\infty f_m\nonumber\\
&=&\sum_{m=0}^\infty{}'\tilde f_m-\frac12\tilde f_0,
\end{eqnarray}
where $\tilde f_m$ is defined by continuity,
\begin{equation}
\tilde f_m=\left\{\begin{array}{cc}
f_m,&m>0,\\
\lim_{m\to0}f_m,&m=0.\end{array}\right.
\end{equation}
Then by using the EM formula,
\begin{eqnarray}
{\mathcal   F}^T&=&\frac{\beta}{2\pi}\int_0^\infty d\zeta\,f(\zeta)+\frac{\zeta(3)}
{8\pi\beta a^3}-\frac{\pi^2}{45}\left(\frac{a}{\beta}\right)^4\nonumber\\
&=&-\frac{\pi^2}{240 a^4}\left[1+\frac{16}{3}\left(\frac{a}{\beta}
\right)^4\right]+\frac{\zeta(3)}{8\pi a^3}T, \quad aT\ll1.
\label{linterm}
\end{eqnarray}
The same result for the low-temperature
limit is extracted through use of the Poisson sum formula, as, for example,
discussed in Ref.~\cite{milton01}.  Let us refer to these results,
with the TE zero mode excluded, as the modified ideal metal model.

Exclusion of the TE zero mode 
 will reduce the linear dependence at high temperature by a factor of
two, but this is not observable by present experiments.  
The main problem, however, is that it adds a linear
term at low temperature, which is given in Eq.~(\ref{linterm}), 
up to exponentially small corrections \cite{milton01}.

 There are apparently two serious problems with the result (\ref{linterm}):
\begin{itemize}
\item  It would seem to be ruled out by experiment.  The ratio of the
linear term to the $T=0$ term is
\begin{subequations}
\begin{equation}
\Delta=\frac{30\zeta(3)}{\pi^3}aT=1.16 aT,
\end{equation}
or putting in the numbers (300 K $ = (38.7)^{-1}$ eV, $\hbar c=197$ MeV fm)
\begin{equation}
\Delta=0.15\left(\frac{T}{300\mbox{ K }}\right)\left(\frac{a}{1 \mu\mbox{m}}
\right),
\end{equation}
\end{subequations}
or as Klimchitskaya observed \cite{klim}, 
there is a 15\% effect at room temperature at
a separation of one micron. One would have expected
 this to have been been seen by Lamoreaux
\cite{lamoreaux97}; his experiment was reported to be in agreement with the 
conventional theoretical prediction at the level of 5\%. 
\item Another  serious problem is the apparent thermodynamic 
inconsistency.  A linear term
in the force implies a linear term in the free energy (per unit area),
\begin{equation}
F=F_0+\frac{\zeta(3)}{16\pi a^2}T,\quad aT\ll1,
\label{linearfe}
\end{equation}
which implies a nonzero contribution to the entropy/area at zero temperature:
\begin{equation}
S=-\left(\frac{\partial F}{\partial T}\right)_V=-\frac{\zeta(3)}{16\pi a^2}.
\label{2.17}
\end{equation}
\end{itemize}
Taken at face value, this statement appears to be incorrect. 
We will discuss this problem more closely in Sec.~\ref{V},
 and will find that although a linear temperature dependence will
occur at room temperature, the entropy will go to zero as the temperature
goes to zero. The point is that the free energy $F$ for a finite 
$\varepsilon$ always will have a zero slope at $T=0$, 
thus ensuring that $S=0$ at $T=0$. The apparent conflict with 
Eq.~(\ref{2.17}) or Eq.~(\ref{linterm}) is due to the fact that the 
curvature of $F(T)$ near $T=0$ becomes infinite when 
$\varepsilon \rightarrow \infty$.
So Eqs.~(\ref{linearfe}) and (\ref{2.17}), corresponding to the modified
ideal metal model, describe real metals approximately only for low, but
not zero temperature---See, for example, Eq.~(\ref{59c}).

\section{Casimir free energy, entropy, and internal energy}
\label{IV}

The Casimir surface force density ${\mathcal   F}^T$ between two dielectric plates
separated by a distance $a$ can be written as
\begin{equation}
{\mathcal   F}^T=-\frac{1}{\pi \beta}{{\sum_{m=0}^\infty }}{}^\prime
\int_{\zeta_m}^\infty q^2dq
\left[\frac{A_me^{-2qa}}{1-A_me^{-2qa}}+\frac{B_me^{-2qa}}{1-
B_me^{-2qa}} \right].
\label{3}
\end{equation}
(We follow the conventions of Ref.~\cite{hoye98} and further references
therein; here we further set $\hbar=c=1$.)  The relation between
$q$ and the transverse wave vector $\bf{k}_\perp$ is $q^2=k_\perp
^2+\zeta_m^2$, where  $\zeta_m=2\pi m/\beta$. 
Furthermore
\begin{subequations}
\begin{eqnarray}
 A_m&=&\left(\frac{\varepsilon p-s}{\varepsilon p+s}\right)^2,\quad
B_m=\left( \frac{s-p}{s+p}\right)^2, \label{4a}\\
s^2&=&\varepsilon -1+p^2, \quad p=\frac{q}{\zeta_m},
\label{4b}
\end{eqnarray}
\end{subequations}
with $\varepsilon(i\zeta_m)$ being the permittivity.
Note that whenever $\varepsilon$ is constant, the
$A_m$ and $B_m$ depend on $m$ and $q$ only in the combination
$p$,
\begin{equation}
A_m(q)=A(p),\quad B_m(q)=B(p).
\end{equation}
(This result may also be found in standard references such as 
Ref.~\cite{milton01}.)

The free energy $F$ per unit area can be obtained from Eq.~(\ref{3}) by
integration with respect to $a$ since ${\mathcal   F}^T=-\partial F/\partial a$.
We get \cite{hoye01}
\begin{subequations}
\begin{equation}
\beta F=\frac{1}{2\pi}{\sum_{m=0}^\infty }{}^\prime
\int_{\zeta_m}^\infty [\ln(1-\lambda^\textrm{TM})+\ln(1-\lambda^\textrm{TE})]q\,dq,
\label{5a}
\end{equation}
where
\begin{equation}
\lambda^\textrm{TM}=A_me^{-2qa},\quad \lambda^\textrm{TE}=B_me^{-2qa}.
\label{5b}
\end{equation}
\end{subequations}
(In the notation of Ref.~\cite{hoye01}, $\lambda_\varepsilon \equiv
\lambda^\textrm{TM},\,\, \lambda \equiv \lambda^\textrm{TE}$.)

{}From thermodynamics the entropy $S$ and internal energy $U$ (both per
unit area) are related to $F$ by $F=U-TS$, implying
\begin{equation}
 S=-\frac{\partial F}{\partial T},\quad \textrm{and thus}\quad 
U=\frac{\partial(\beta F)}{\partial \beta}. 
\label{star}
\end{equation}
As mentioned above the behaviour of $S$ as $T\rightarrow 0$ has been
disputed, especially for metals where $ \varepsilon \rightarrow \infty$.
We now see the mathematical root of the problem: The quantities
$A_m=B_m\rightarrow 1$ in the  $\varepsilon\to\infty$ 
 limit except that $B_0=0$ for any
finite $\varepsilon$. So the question has been whether $B_0=0$ or
$B_0=1$ or something in between should be used in this limit as
results will differ for finite $T$, producing,
 as we saw above, a difference in the force
linear in
$T$. The corresponding difference in entropy will thus be  nonzero.
Such a difference would lead to a violation of the third law
of thermodynamics, which states that the entropy of a 
system with a nondegenerate ground state should be zero at $T=0$. 
Inclusion of the interaction between the plates at different separations
cannot change this general
property.
 We will show that this discrepancy vanishes when the limit
$\varepsilon \rightarrow \infty$ is considered carefully, by using the
Euler-Maclaurin summation formula.  Also, we will perform explicit
analytic evaluation for any $T$ for metallic plates in the case where
$\varepsilon \rightarrow \infty$ for all $\zeta$.

We will consider this latter case first. It is the case of ``ideal"
metals mentioned in Sec.~\ref{I}  and already considered 
briefly in Sec.~\ref{II}.

\subsection{``Ideal" metals}
\label{IV.1}
With $\varepsilon =\infty$ we have $A_m=B_m=1$ where we now also
put $B_0=1$, i.e.,
$\lambda^\textrm{TM}=\lambda^\textrm{TE}=e^{-2qa}$. To remove the
$\zeta$-dependence in the lower limit of integration in Eq.~(\ref{5a}), 
it is convenient 
to use the quantity $p$ of Eq.~(\ref{4b}) as a new variable. Expanding the
logarithmic terms in Eq.~(\ref{5a}) and keeping only the leading term, we get the
task of calculating
\begin{equation}
F\approx -\frac1{2\pi\beta} I_1,\quad I_1\equiv
2{\sum_{m=0}^\infty}{}^\prime\zeta_m^2\int_1^\infty
 p\,e^{-2\gamma mp}\,dp,
\label{7}
\end{equation}
where
\begin{equation}
\gamma m=a\zeta_m=\frac{2\pi a}{\beta}m.
\label{8}
\end{equation}
Carrying out the integration in Eq.~(\ref{7}) we obtain
\begin{subequations}
\begin{equation}
I_1=\frac{1}{(2a)^2}\,2{\sum_{m=0}^\infty }{}^\prime L_m,
\label{9}
\end{equation}
with
\begin{equation}
L_m=(2\gamma m+1)e^{-2\gamma m}.
\label{10}
\end{equation}
(It is easy to check that this result is correct at $m=0$, where $p$ is not
defined.)
\end{subequations}
We encounter the following sums
\begin{subequations}
\label{11}
\begin{eqnarray}
 s_0(\gamma)&=&2{\sum_{m=0}^\infty}{}^\prime e^{-2\gamma m}=\coth \gamma, 
\label{11a}\\
 s_k(\gamma)&=&2\sum_{m=0}^\infty(2\gamma m)^ke^{-2m\gamma}
=(-\gamma)^k\frac{\partial ^k s_0}{\partial
\gamma^k},\label{11b}
\end{eqnarray}
\end{subequations}
so that
\begin{subequations}
\begin{eqnarray}
 s_1&=&\frac{\gamma}{\sinh^2 \gamma},\label{11c}\\
s_2&=&\frac{2\gamma^2\cosh \gamma}{\sinh^3 \gamma}, \label{11d}\\
s_3&=&\gamma^3\,\frac{6+4\sinh^2\gamma}{\sinh^4\gamma}.
\label{11e}
\end{eqnarray}
\end{subequations}
The quantity $I_1$ is given by the first two of these sums,
\begin{equation}
I_1(\gamma)=\frac{1}{(2a)^2}\,[s_1(\gamma)+s_0(\gamma)].
\label{13}
\end{equation}
Alternatively, one could just first perform the summation in Eq.~(\ref{7}) (for $m \ge 1$) 
and then
integrate. This summation yields $s_2(\gamma p)$. By
subsequently integrating $s_2$ by parts the quantity $(s_1+s_0)$ in
Eq.~(\ref{13}) is recovered (adding the $m=0$ term separately).

By further expansion of the logarithm in Eq.~(\ref{5a}) one obtains terms
$\lambda^k/k$ to be integrated and summed like Eq.~(\ref{7}). Performing the
same steps as before, we find that the result (\ref{13}) generalizes to
\begin{equation}
F=-\frac{1}{8\pi \beta a^2}\,\sum_{k=1}^\infty \frac{1}{k^3}[
s_1(\gamma k)+s_0(\gamma k)],
\label{15}
\end{equation}
valid for arbitrary temperature.

The surface force per area (\ref{3}) can now be obtained via ${\mathcal   F}^T
=-\partial
F/\partial a$ utilizing $\gamma \propto a$ [Eq.~(\ref{8})]. This yields
\begin{equation}
{\mathcal   F}^T=-\frac{1}{8\pi \beta a^3}\sum_{k=1}^\infty \frac{1}{k^3}[
s_2(\gamma k)+2s_1(\gamma k)+2s_0(\gamma k)].
\label{16}
\end{equation}
The same result is also obtained by evaluating expression (\ref{3})
(with $A_m=B_m=1$) in the
same way as expression (\ref{5a}) for $F$ was evaluated above. Using the
second method, mentioned below Eq.~(\ref{13}), one finds that the integration
of $s_3(\gamma p)/p$ yields the combination of $s_i$ present in Eq.~(\ref{16}).

Considering the $T\rightarrow 0$ limit, which implies the $\gamma
\rightarrow 0$ limit, one obtains
\begin{equation}
{\mathcal   F}^T=-\frac{1}{8\pi \beta a^3}\sum_{k=1}^\infty
\frac{1}{k^3}\,\frac{6}{\gamma k}=-\frac{\pi^2}{240 a^4},
\label{17}
\end{equation}
using the limiting values of expressions 
(\ref{11a}), (\ref{11c}), and (\ref{11d}).
This is the well known Casimir result for idealized metallic plates at
$T=0$, seen in Eq.~(\ref{linterm}).

The internal energy $U$ is now found from Eqs.~(\ref{star}), (\ref{8}), 
and (\ref{15}) to be
\begin{equation}
U=-\gamma^2\, \,\frac{\partial (F/\gamma)}{\partial \gamma}=
-\frac{1}{8\pi \beta a^2}\sum_{k=1}^\infty \frac{1}{k^3}\,s_2(\gamma
k),\label{19}
\end{equation}
and similarly an expression for the entropy $S$ follows from
\begin{equation}
S=-{2\pi a}\frac{\partial F}{\partial
\gamma}=\frac{U-F}{T}=-\frac1{8\pi a^2}\sum_{k=1}^\infty\frac1{k^3}[s_2(\gamma k)
-s_1(\gamma k)-s_0(\gamma k)].
\label{20}
\end{equation}
with Eqs.~(\ref{15}) and (\ref{19}) inserted.

Now we can analyze the thermodynamic quantities
in the low temperature limit using the properties of
$s_k$ as defined by  Eqs.~(\ref{11a})--(\ref{11e}). 
We have for low temperature,\footnote{Actually, for a room-temperature
experiment, $\gamma$ need not be small.  For $T=300$ K and $a=1 \mu$m,
$\gamma=0.823$.} where $\gamma \propto T\rightarrow
0$ 
\begin{subequations}
\label{21}
\begin{eqnarray}
s_0&=&\frac{1}{\gamma}+\frac{1}{3}\gamma-\frac{1}{45}\gamma^3+\dots,
 \label{21a}\\
s_1&=&\frac{1}{\gamma}-\frac{1}{3}\gamma+\frac{1}{15}\gamma^3-\dots,
\label{21b}\\
s_2&=&\frac{2}{\gamma}-\frac{2}{15}\gamma^3+\dots, 
 \label{21c}\\
s_3&=&\frac{6}{\gamma}+\frac{2}{15}\gamma^3-\dots.\label{21d}
\end{eqnarray}
\end{subequations}
Inserting this into expressions (\ref{15}), (\ref{16}) or (\ref{19}) 
one finds that the
terms linear in $\gamma$ vanish.\footnote{This is actually stronger
than necessary to insure vanishing entropy, since such terms would give
$T^2$ terms in the energy or free energy.}
 Thus the entropy (\ref{20}) vanishes, as it
should in accordance with the third law of thermodynamics.

To obtain the leading correction to the $T=0$ result for finite $T$
one must consider the $\gamma^3$ term in the power series expansion of the 
summand in Eq.~(\ref{16}).
 However, the summation of this term with
respect to $k$ diverges,\footnote{For this reason, the alternate expression
(3.35) in Ref.~\cite{milton01} might be preferred.  See Eq.~(\ref{42}) below.}
because the expansion of $s_n(\gamma k)$ is not valid for large $k$.
 For small $\gamma$ one can instead integrate, without expanding,
using the Euler-Maclaurin summation formula  (\ref{22})
 to obtain a finite
correction to the zero-temperature result.  Using
Eq.~(\ref{22}) to evaluate expression (\ref{16}), the $\gamma \rightarrow 0$
expression (\ref{17}) has to be subtracted to make $f(0)$ finite. Putting
$x=\gamma k$ we have, apart from a prefactor,
\begin{equation}
f(x)=\frac{1}{x^3}\left[s_2(x)+2s_1(x)+2s_0(x)-\frac{6}{x}\right],
\label{23}
\end{equation}
with $f(0)=-2/45$ in view of the expansions  (\ref{21a})--(\ref{21c}).
Integrating and using
expressions  (\ref{11a}), (\ref{11c}), (\ref{11d}),  we obtain
\begin{equation}
\int_0^\infty f(x)\,dx=-\frac{1}{x^2}\left[s_1(x)+s_0(x)-\frac{2}{x}\right]
{\bigg|}_0^\infty
=0.
\label{24}
\end{equation}
Including the $T=0$ result (\ref{17}) we thus find
\begin{eqnarray}
{\mathcal   F}^T&=&-\frac{1}{8\pi \beta
a^3}\left[\frac{6}{\gamma}\frac{\pi^4}{90}-\frac{1}{2}f(0)\gamma^3
\right]\nonumber\\
&=&-\frac{\pi^2}{240
a^4}\left[1+\frac{1}{3}\left(\frac{2a}{\beta}\right)^4\right],\quad aT\ll1,
\label{25}
\end{eqnarray}
where we have inserted expression (\ref{8}) for $\gamma$ and noted that there
is no $k=0$ term in Eq.~(\ref{16}), i.e., $f(0)$ is to be subtracted from
expression (\ref{22}).  All the odd derivatives in the Euler-Maclaurin formula
vanish because $f(x)$ is even.
 It should be noted that the expression for ${\mathcal   F}^T$ is in
agreement with what has been found earlier [cf.~Eq.~(\ref{linterm})], 
via alternative methods, by
Milton \cite{milton01}, Klimchitskaya and Mostepanenko
\cite{klimchitskaya01}, Sauer \cite{sauer62}, Mehra \cite{mehra67}, and
others, where the exponentially small correction to the above formula is
also given.

The free energy (\ref{15}) can be obtained from ${\mathcal   F}^T=-\partial F/\partial
a$, but this leaves a temperature dependent constant of integration. So
instead we make use of the method above, where from Eq.~(\ref{15})
\begin{equation}
f(x)=\frac{1}{x^3}\left[s_1(x)+s_0(x)-\frac{2}{x}\right],
\label{26}
\end{equation}
and where now $f(0)=2/45$. With Eq.~(\ref{26}) we get a nonzero integral
\begin{eqnarray}
C&=&\int_0^\infty
f(x)\,dx=-\int_0^\infty\frac{1}{x}\,\frac{d}{dx}\left(\frac{1}{x}\coth
x-\frac{1}{3}-\frac{1}{x^2}\right)dx \nonumber\\
&=&\int_0^\infty\frac{1}{x^3}\left(\frac{1}{x}+\frac{x}{3}-\coth
x\right) dx,
\label{27}
\end{eqnarray}
using partial integration.
The integral (\ref{27}) may be easily evaluated by contour methods.
Due to symmetry the integral can be extended to minus infinity and then the
contour of integration can be distorted into one which encircles the poles
along the positive imaginary axis.
Since $\coth z$ has poles at $z=i\pi m$ with $m$ integer we 
get\footnote{This low temperature $T^3$ dependence in $F$, which does not 
contribute to the force, is determined by the linear high temperature
behavior of ${\mathcal   F}^T$---see Ref.~\cite{milton01}, Sec.~3.2.1.}
\begin{equation}
C=\frac{1}{2}2\pi i \sum_{m=1}^\infty \frac{-1}{(\pi
im)^3}=\frac{1}{\pi^2}\zeta(3).
\label{46}
\end{equation}

In view of this  result as well as Eq.~(\ref{17})
  we obtain for the free energy ($dk=dx/\gamma$)
\begin{eqnarray}
 F&=&-\frac{1}{8\pi \beta a^2}\left(
\frac{2}{\gamma}\frac{\pi^4}{90}+\gamma^3\left(\frac{C}{\gamma}-\frac{1}{2}
f(0)\right)\right) \nonumber\\
&=&-\frac{\pi^2}{720a^3}\left(1+45\left(\frac{2a}\beta\right)^3\frac{\zeta(3)}
{\pi^3}-\left(\frac{2a}\beta\right)^4\right),\quad aT\ll1.
\label{28}
\end{eqnarray}
This result, including its exponentially small correction, is given in 
Ref.~\cite{milton01} and references therein.
The internal energy $U$, which can be most easily be evaluated using
Eq.~(\ref{star}), can also be computed by the method above, starting from
the sum (\ref{19}). Then
\begin{equation}
f(x)=\frac{1}{x^3}\left(s_2(x)-\frac{2}{x}\right)=-\frac{1}{x^2}
\frac{d}{dx}\left(s_1(x)+s_0(x)-\frac{2}{x}\right),
\label{29}
\end{equation}
with $f(0)=-2/15$. Partial integration replaces the $C$ of Eq.~(\ref{27})
with $-2C$, and we obtain
\begin{equation}
U=-\frac{\pi^2}{720 a^3}\left[1-90\left(\frac{2a}\beta\right)^3\frac{\zeta(3)}
{\pi^3}+3\left(\frac{2a}\beta\right)^4\right],\quad aT\ll1.
\label{30}
\end{equation}
With Eq.~(\ref{20}) the entropy thus becomes (recall that $B_0=1$ is assumed)
\begin{equation}
S=\frac{U-F}{T}\sim \frac{3\zeta(3)}{2\pi}\,T^2-\frac{4\pi^2a}{45}T^3,
\quad aT\ll1.
\label{31}
\end{equation}

\subsection{Equivalence with earlier results}
\label{IV.2}
Equivalence with previous derivations can be shown for any $\gamma$. It
is then convenient to utilize the Poisson summation   formula. If
$\tilde{c}(k)$ is the Fourier transform of $c(x)$,
defined by
\begin{equation}
\tilde c(k)=\int_{-\infty}^\infty dx\,c(x)\,e^{ikx},
\end{equation}
then
\begin{equation}
\sum_{n=-\infty}^\infty c(n)=\sum_{m=-\infty}^\infty
\tilde{c}(2\pi m).
\label{32}
\end{equation}
With $c(x)=e^{-2\gamma |x|}$ one finds
\begin{equation}
\tilde{c}(2\pi m)=\int_{-\infty}^\infty e^{-2\gamma |x|+2\pi
mxi}\,dx=\frac{\gamma}{\gamma^2+(\pi m)^2}.
\label{33}
\end{equation}
Thus
\begin{equation}
\sum_{m=-\infty}^\infty \frac{\gamma}{\gamma^2+(\pi
m)^2}=\sum_{n=-\infty}^\infty e^{-2\gamma |n|}=\coth \gamma,
\label{34}
\end{equation}
the familiar cotangent expansion,
which can be verified in many different ways (cf.~Ref.~\cite{hoye81}).

In Eqs.~(\ref{15}) and (\ref{16}) one of the sums is [$s_0(x)=\coth x$]
\begin{subequations}
\begin{equation}
S_0=\sum_{k=1}^\infty \frac{1}{k^3}\coth (\gamma
k)=\sum_{k=1}^\infty \sum_{m=-\infty}^\infty S_{0mk},
\label{35}
\end{equation}
where with Eq.~(\ref{34})
\begin{eqnarray}
S_{0mk}=\frac{\gamma k}{k^3[(\gamma k)^2+(\pi m)^2]}
=\frac{1}{mu}\left[\frac{1}{k^2}-\frac{1}{k^2+(u/\pi)^2}\right],\quad
u=\pi^2m/\gamma.
\label{36}
\end{eqnarray}
\end{subequations}
Summation first with respect to $k$ where also the result (\ref{34}) is
utilized then gives
\begin{equation}
S_{0m}=\sum_{k=1}^\infty
S_{0mk}=\frac{1}{mu}\left[\frac{\pi^2}{6}-\frac{\pi^2}{2u}\left(\coth
u-\frac{1}{u}\right)\right].
\label{37}
\end{equation}
In the limit $\gamma \rightarrow 0$ only the $m=0$ term remains, and
we get the $T=0$ result if we use the expansion (\ref{21a})  ($u\rightarrow
0$)
\begin{equation}
S_{00}\rightarrow
\frac{1}{mu}\left(-\frac{\pi^2}{2u}\right)\left(-\frac{u^3}{45}\right)
=\frac{\pi^4}{90}\frac{1}{\gamma},
\label{38}
\end{equation}
which is consistent with the $1/k^4$ sum occurring in Eq.~(\ref{17}).

To obtain the free energy $F$ and the force ${\mathcal   F}^T$ there are sums $S_1$
and $S_2$ that follow from the $s_1$ and $s_2$ of Eqs.~(\ref{11c})
and (\ref{11d}).  And like
Eqs.~(\ref{35}) the relations between the various $s_i$ lead to
\begin{equation}
S_{1m}=-\gamma\,\frac{\partial}{\partial \gamma}g=ug',
\label{39}
\end{equation}
where $g(u)=S_{0m}$. Also:
\begin{equation}
S_{2m}=\gamma^2\frac{\partial}{\partial \gamma}\left(
-\frac{u}{\gamma}g'\right)=2ug'+u^2g''.
\label{40}
\end{equation}
So to obtain ${\mathcal   F}^T$ we need, because
\begin{equation}
\left(2+4u\frac{\partial}{\partial u}+u^2\frac{\partial^2}{\partial u^2}\right)
\frac1{u^2}g(u)=g''(u),
\end{equation}
the combination
\begin{eqnarray}
S_{2m}+2S_{1m}+2S_{0m}&=&\frac{\pi^2}{6m}\,\frac{d^2}{du^2}\left[u-3(\coth
 u-\frac{1}{u})\right] \nonumber\\
&=&\frac{\pi^2}{m}\left(\frac{1}{u^3}-\frac{\cosh u}{\sinh^3
u}\right)\stackrel{m\rightarrow 0}{\longrightarrow
}\frac{\pi^2}{m}\frac{u}{15}=\frac{\pi^4}{15}\frac{1}{\gamma}.
\label{41}
\end{eqnarray}
Altogether, restricting $m$ to positive values due to symmetry, the
expression (\ref{16}) can be reexpressed as
($u=\pi^2m/\gamma,~\gamma=2\pi a/\beta$)
\begin{equation}
{\mathcal   F}^T=-\frac{\pi^2}{240a^4}\left[1+30\sum_{m=1}^\infty
\left(\frac{1}{u^4}-\frac{\cosh u}{u\sinh^3u}\right)\right],
\label{42}
\end{equation}
which is the desired known expression. (For example, compare Eq.~(3.35) of
Ref.~\cite{milton01}.)

To calculate the free energy (\ref{15}) one likewise needs
\begin{eqnarray}
S_{1m}+S_{0m}&=&\frac{\pi^2}{6m}\frac{d}{du}\left[1-3\left(\frac{\coth
u}{u}-\frac{1}{u^2}\right) \right]\nonumber\\
&=&\frac{\pi^2}{2m}\left[\frac{\coth
u}{u^2}+\frac{1}{u\sinh^2u}-\frac{2}{u^3}\right]
 \nonumber\\
& \stackrel{m\rightarrow 0}{\longrightarrow}&
\frac{\pi^2}{2m}\left(-\frac{1}{45}+\frac{1}{15}\right)u=\frac{\pi^4}{
45}\frac{1}{\gamma}.
\label{43}
\end{eqnarray}
Thus the free energy becomes
\begin{equation}
F=-\frac{\pi^2}{720
a^3}\left[1+45\sum_{m=1}^\infty\left(\frac{\coth
u}{u^3}+\frac{1}{u^2\sinh^2 u}-\frac{2}{u^4}\right) \right].
\label{44}
\end{equation}
Compared with the small $T$ or $\gamma$ expansion (\ref{28}) it is clear that
the last term of Eq.~(\ref{44}) gives the $T^4 = \beta^{-4}$ term of
(\ref{28}). The coefficient $C$ can also be identified from Eq.~(\ref{44}). As
$\coth u\rightarrow 1$ when $\gamma \rightarrow 0$ we must have, when
comparing with Eq.~(\ref{28}),
\begin{subequations}
\begin{equation}
\left(\frac{\gamma}{\pi}\right)^4\frac{45
C}{\gamma}=45\sum_{m=1}^\infty
\frac{1}{u^3}=45\left(\frac{\gamma}{\pi^2}\right)^3\sum_{m=1}^\infty
 \frac{1}{m^3}, \end{equation}
or
\begin{equation}
C=\frac{1}{\pi^2}\zeta(3),
\label{45}
\end{equation}
\end{subequations}
which is in agreement with Eq.~(\ref{46}).

\section{Finite permittivity. Real metals}
\label{V}
\subsection{Two harmonic oscillator models}
\label{V.1}
With finite permittivity $\varepsilon$ the $A_m$ and $B_m$ of Eq.~(\ref{4a})
will vary with $p$. Especially $B_m \rightarrow 0$ as $p\rightarrow
\infty$ or $\zeta_m \rightarrow 0$ ($\zeta_m=2\pi m/\beta$). In
the high temperature or classical limit only the Matsubara frequency
$\zeta =0$ (or $m=0$) can contribute as $\beta \rightarrow 0$.
Thus, in the classical limit one has the result that the TE mode does
not contribute at all. Physically, this means that the temperature
becomes so high that only the static dipole-dipole interaction
contributes (the $\zeta \rightarrow 0$ limit of the TM mode). In our
opinion this somewhat unexpected behaviour is related to the peculiar
type of interaction that exists between the canonical momentum $\mathbf  p$
of a particle 
and the electromagnetic vector potential $\mathbf{  A(r},t)$, which for a
particle of mass $m$ and charge $q$ 
is $(\mathbf{  p}-q\mathbf{  A})^2/2m$. In addition to the standard
cross term interaction $\mathbf{  p\cdot A}$ this also implies an interaction
$\mathbf{  A}^2$.

As an illustration of the above we can consider two models, in each
of which two harmonic
oscillators interact via a third one. These oscillators represent
a simplified picture of our polarizable parallel plates interacting via
the electromagnetic field. The classical partition function of a
harmonic oscillator with frequency $\omega$ is const/$(\beta \omega)
\sim 1/\sqrt{\omega^2}$, which gives a free energy $\sim
\ln(\omega^2)$. Thus for three noninteracting harmonic oscillators the
inverse partition function is proportional to $\sqrt{Q}$, where
\begin{subequations}
\begin{equation}
Q=a_1a_2a_3,
\label{47}
\end{equation}
with
\begin{equation}
 a_i=\omega_i^2,\quad (i=1,2,3). \end{equation}
\end{subequations}
(The quantity $a_3$ corresponds to $k_\perp^2$ above.)
By quantization using the path integral method \cite{hoye81,brevik88}, the classical system is
split into a set of harmonic oscillator
systems described by Matsubara frequencies.
Expression (\ref{47}) is replaced by
\begin{subequations}
\begin{equation}
Q=A_1A_2A_3,
\label{48}
\end{equation}
where
\begin{equation}
 A_i=\omega_i^2+\zeta^2=a_i+\zeta^2.\end{equation}
\end{subequations}
(For real frequencies, $\omega=i\zeta$, $1/A_i$ determines the response
to an external oscillating force acting on the oscillator.)

Now add interactions, of strength proportional to $c$, between the
third oscillator and the other two. The usual form of this interaction is
$c x_ix_j$, where $x_i$ and $x_j$ are coordinates.  Let this
constitute the first model, which is analogous to the TM mode.
Then the quantity $Q$ becomes the
determinant of the matrix,
\begin{subequations}
\begin{eqnarray}
 Q&=&\left| \begin{array}{ccc}
A_1  & 0  &  c\\
0    &A_2 &  c\\
c    & c  & A_3
\end{array}
\right| =A_1A_2A_3-c^2(A_1+A_2)\nonumber\\
&=&A_1A_2A_3(1-D_1)(1-D_2)\left(1-\frac{D_1D_2}{(1-D_1)(1-D_2)}\right),
\label{49}
\end{eqnarray}
where
\begin{equation}
 D_i=\frac{1}{A_i}\frac{c^2}{A_3}\quad (i=1,2).\end{equation}
\end{subequations}
 The quantum free energy for this system of three coupled oscillators
 is given by summing over the Matsubara frequencies, as in Eq.~(\ref{5a}):
 \begin{equation}
 \beta F= \frac12\lim_{N \rightarrow \infty}\sum_{m=1}^N (\ln Q(\zeta_m )
 +3\ln \eta^2),
\label{49a}
 \end{equation}
where $\eta=\beta/N$ and $\zeta^2$ is replaced by 
$2(1-\cos (\zeta \eta))/\eta^2\; (=\zeta^2 +...)$ in the $A_1A_2A_3$ term 
of Eq.~(\ref{49}). The limiting procedure $N\rightarrow \infty$ is 
required to make the full free energy well defined. This means that the 
path integral representation of a harmonic oscillator is discretized by 
dividing the imaginary time of periodicity $\beta$ into $N$ pieces each 
of length $\eta$ as done in Ref.~\cite{hoye81}. There, in an appendix an 
explicit evaluation was performed for one single oscillator.

The various factors in Eq.~(\ref{49}) can be interpreted as follows: The 
product $A_1 A_2 A_3$ corresponds to the
noninteracting system, the next two factors represent the result of interaction
of single oscillators with the third one, while the last one is the
contribution from the induced interaction between the two single
oscillators via the third one. The logarithm of the last term is the
analogue of the Casimir free energy. In this respect the term $c^2/A_3$
represents the induced interaction. Furthermore the $1/a_i$ ($i=1,2$)
represents the ``bare" polarizability of noninteracting particles which
for nonzero $\zeta$ becomes $1/A_i$. Due to interaction with the
``radiation" field this polarizability is modified into
$1/(A_i(1-D_i))$ ($i=1,2$), where $D_i$ represents a ``radiation"
reaction from the ``field" upon each single oscillator.

The above represents the ordinary situation, analogous to
the TM mode.  To model the TE mode, we can consider an
analogy with the electromagnetic interaction in which the third oscillator
can interact with the momenta of the first two. The analogous
interaction will be $(p_i-\textrm{const.}\, x_3)^2/2m_i\quad (i=1,2;~ m_i$ is
mass), including the unperturbed $p_i^2$ term. By evaluation of the
classical partition function one now finds that the interaction from
const.\,$x_3$ has no influence. (This is the analogue of classical
diamagnetism which is equal to zero, as const.\,$x_3$ is seen to have no
influence on the result when $p_i$ is integrated first.)

Quantum mechanically, the problem is a bit more complex. However, we
can now exchange the roles of momenta and coordinates of the first two
oscillators, i.e., we introduce a momentum representation. Then the
interaction with the third oscillator can be written as ($i=1,2$)
\begin{equation}
\textrm{const.}\,a_i\left(x_i-\frac{c}{a_i}x_3\right)^2=\textrm{const.}\left(a_i
x_i^2-2cx_i x_3+\frac{c^2}{a_i}x_3^2 \right).
\label{50}
\end{equation}
Now the last quadratic term adds to the energy of the third oscillator
alone. Thus, compared to the first model considered above, $a_3$ is changed
while the other $a_i$ remain unchanged:
\begin{subequations}
\begin{equation}
a_3 \rightarrow a_3+c^2/a_1+c^2/a_2.
\label{51}
\end{equation}
Likewise in the quantum case
\begin{equation}
A_3\rightarrow A_3+c^2/a_1+c^2/a_2.
\label{52}
\end{equation}
\end{subequations}
The quantity $Q$  can still be written in
the form (\ref{49}), but due to the change of $a_3$,
 the $(1/A_i)$ ($i=1,2$) is replaced by
$1/A_i-1/a_i=-\zeta^2/(a_iA_i)$ when evaluating $D_i$, i.e.,
\begin{equation}
D_i=-\frac{\zeta^2}{a_i(a_i+\zeta^2)}\frac{c^2}{A_3}.
\label{53}
\end{equation}
 The induced 
(analogous to the Casimir) free energy is again given by the logarithm of the 
third term in Eq.~(\ref{49}). At zero and finite temperatures the latter 
logarithm is negative, and the free energy
\begin{equation}
\frac{T}2\sum_{m= -\infty}^\infty \ln 
\left( 1-\frac{D_1D_2}{(1-D_1)(1-D_2)}\right)
\end{equation}
is negative. Note that here the limiting procedure of Eq.~(\ref{49a}) 
is not needed as sums for free energy differences converge, without 
difficulties.  In the classical 
limit, however, the induced free energy becomes equal to zero 
($D_i \rightarrow 0$ implies that we get the logarithm of unity). We note 
the analogy: At high temperatures the same is true for the TE mode in the
Casimir effect.
There exists thus at least somewhere a finite temperature interval for which 
the Casimir free energy \textit{increases} with increasing temperature. In turn, 
this means that the Casimir entropy $S=-\partial F/\partial T$ becomes 
negative in this interval.

This is a counterintuitive effect, but is physically due to the fact that we 
are dealing with the induced interaction part of the free energy of a 
composite system. We cannot apply usual thermodynamic restrictions such as 
positiveness of entropy to a ``subsystem" of this sort. There exists actually 
a striking analogy with the peculiar formal properties one encounters in 
connection with the theory of the electromagnetic field in a continuous 
medium. The electromagnetic energy-momentum tensor that experimentally turns 
out to be definitely the best alternative when dealing with high-frequency 
effects, is the Minkowski tensor (cf., for instance, Ref.~\cite{brevik79}). 
This tensor is however nonsymmetric, apparently breaking general conservation 
principles for angular momentum. The reason why this peculiar behaviour is yet 
quite legitimate physically, is that phenomenological electrodynamic theory is 
dealing only with a subsystem (the field itself plus its interaction with 
matter), and we cannot apply the same formal restrictions on it as we could 
if the system were closed.

\subsection{Real metal}
\label{V.2}

In the limit of an ideal metal ($\varepsilon \rightarrow \infty $) the 
traditional (SDM) prescription, as mentioned in the Introduction, implies
that $ A_m =B_m =1$ for all $m$. In addition, as also mentioned previously, 
thermodynamic arguments have been given, claiming that the entropy does not 
become zero at $T=0$ in violation of the third law of thermodynamics 
if $B_0=0$ is used \cite{bezerra02a}.
However, we do not find this to be the case; as we will show below, the entropy
will be zero as required at $T=0$, even for a metal that is not idealized and 
where one bases the analysis on
the value $B_0=0$.

Let us go back to Eq.~(\ref{7}). That equation was obtained by expanding
Eq.~(\ref{5a}) to first order in $\lambda$ under the assumption that $A_m
=B_m=1$.  Doing the same expansion
for finite permittivity, we obtain an integrand
which contains a term with
a factor $B_m$ (or $A_m$) that varies with $p=q/\zeta_m$ such that
$B_m \rightarrow  0$ when $p \rightarrow \infty$. Expanding Eq.~(\ref{5a}) to
higher order one obtains likewise powers of $B_m$ which, because $B_m <1$,
become less important as compared to the case of an ideal metal
(where $B_m=1$). One can first consider the case where $\varepsilon$ is
independent of $\zeta$. When $\varepsilon$ is large one can use as a rough
approximation
\begin{eqnarray}
B_m=\left\{ \begin{array}{ll}
1,  & p<\sqrt{\varepsilon},\\
0,  & p>\sqrt{\varepsilon}.
\end{array}
\right.
\label{54}
\end{eqnarray}
This simple expression for $B_m$ is intended to show essential features that 
will be obtained more accurately in a detailed numerical calculation. 
With this, Eq.~(\ref{11a}) (neglecting the influence of $A_m$) will turn into
\begin{subequations}
\begin{equation}
s_0(\gamma)\to s_0(\gamma)-s_0(\sqrt{\varepsilon}\gamma)
=\coth \gamma -\coth \gamma_c,
\label{55}
\end{equation}
with similar modifications for $s_i~(i=1,2,3)$. Here
\begin{equation}
\gamma_c=\gamma \sqrt{\varepsilon}
\label{56}
\end{equation}
\end{subequations}
is an effective sharp cutoff limit for the integral, 
a crude model for what should be a gradual cutoff for the integral of interest.
[A gradual cutoff will only modify the last term of (\ref{55})
into a sum or integral over terms with varying $\gamma_c$.
Namely, with varying $B=B(p)$, Eq.~(\ref{7}), if we recall the comment below 
Eq.~(\ref{13}), changes into ($B(1)\approx1$ for $\varepsilon$ large)
\begin{eqnarray}
I_1 &=& \frac1{(2a)^2}\int_1^\infty B(p)s_2(\gamma p) \frac{dp}p \nonumber\\
&=& \frac1{(2a)^2}\left[s_0(\gamma)+s_1(\gamma)\right]
+\frac1{(2a)^2}\int_1^\infty[s_0(\gamma p)+s_1(\gamma p)] B'(p)\,dp,
\end{eqnarray}
using partial integration.  
The approximation (\ref{54}) means that $B'(p)=-\delta (p-\sqrt{\varepsilon}).$]

As we did to obtain Eq.~(\ref{31}),
we carry out the sum over $k$ in Eq.~(\ref{20}) while assuming
$\varepsilon$ sufficiently large such that approximation (\ref{54})
can be used. Then as in Eq.~(\ref{55}) one obtains the previous result
minus a term with $\gamma \rightarrow \gamma_c$.
Keeping only the leading term, Eq.~(\ref{31}) is modified into
\begin{equation}
S^{\textrm{TE}}\sim \frac{3\zeta(3)}{4\pi}(1-\varepsilon) T^2,\quad\sqrt{\varepsilon}
aT\ll1.
\label{59a}
\end{equation}
[However, to be more accurate 
$B_m=((\sqrt{\varepsilon}-1)/(\sqrt{\varepsilon}+1))^2$ for
 $p=1$ and thus $B_m<1$ for $p< \sqrt{\varepsilon}$. 
 When this is taken into account, we find that 
 $S^\textrm{TE} \propto -a\varepsilon^{5/2}T^3$ in a more narrow 
region, $\varepsilon^{3/2}\,aT \ll 1$, but that Eq.~(\ref{59a})
holds for $\varepsilon^{-3/2}\ll aT\ll\varepsilon^{-1/2}$.]

Thus the entropy approaches zero as the temperature goes to zero. 
As $\varepsilon$ increases the $T$-dependence becomes more singular,
because the region in which Eq.~(\ref{59a}) is valid becomes more and more
narrow, but the value at $T=0$ stays fixed at zero also in the limit
$\varepsilon\rightarrow\infty$. This contrasts the ideal metal result 
(\ref{2.17}) where $\varepsilon=\infty$ is used.

Again, we note the counterintuitive negative contribution from the TE mode. 
As mentioned earlier, this does not violate the laws of thermodynamics and 
can be understood in terms of the oscillator model analysed in some detail 
in Sec.~\ref{V.1}. Only the total entropy has to increase with increasing 
temperature. And this is the case for the inverse partition function 
(\ref{49}) which represents three interacting harmonic oscillators where the 
$D_i$ are given by Eq.~(\ref{53}). Although the induced entropy becomes 
negative at least in some finite temperature region the total entropy 
will behave properly, as the total system can be decomposed into three 
independent harmonic
oscillators represented by the eigenvalues of the matrix (\ref{49}) with
$A_i$ replaced by $a_i$ ($i=1,2$), and furthermore $A_3$ replaced by the 
right hand side of Eq.~(\ref{51}).

With the simplification (\ref{54}) for the TE-mode the free energy can be 
easily expressed in terms of the ``ideal" metal case analysed in 
Sec.~\ref{IV.1}. Let the ``ideal" metal free energy be $F=F_I(T)$.
{}From Eq.~(\ref{8}) $\gamma\propto T$.
Now the magnification of $\gamma$ to $\gamma_c$ as in Eq.~(\ref{56}) and 
insertion of it in Eq.~(\ref{15})
will change the corresponding free energy to
$(\gamma/\gamma_c)F_I(T\gamma_c/\gamma)=F_I(\sqrt{\varepsilon} T)/\sqrt{\varepsilon}$.
The TM- and TE-modes both contribute the same amounts to (\ref{15}). 
Thus with Eq.~(\ref{54}) the free energy will be
\begin{equation}
F=F(T)=F_I(T)-\frac{1}{2\sqrt{\varepsilon}} F_I(\sqrt{\varepsilon}T).
\label{59b}
\end{equation}
{}From this we have (keeping in each case only the leading temperature
correction)
\begin{eqnarray} 
F(T)=\left\{\begin{array}{ll}
(1- 1/(2\sqrt{\varepsilon}))F_I(0)-
\frac{\zeta(3)}{4\pi}(2-\varepsilon)T^3,
\quad&0\leq aT\ll 1/\sqrt{\varepsilon},\\
F_I(0)+K_I T/2, &1/\sqrt{\varepsilon} \ll aT\ll 1,\\
-K_I T/2, &1\ll aT.
\end{array}
\right.
\label{59c}
\end{eqnarray}
where the constant $K_I=\zeta(3)/(8\pi a^2)$ 
is the magnitude of the slope of the linear dependence of the high 
temperature result of the the ``ideal" metal ($F_I(0)=-\frac{\pi^2}{720a^3}
< 0$). Thus for high
temperatures non-ideal or realistic metals yield one half of the 
``ideal" metal result.   The intermediate form, which holds at room
temperature, is the same as seen in Eq.~(\ref{linearfe}). 
Again, we see that in the $\sqrt{\varepsilon}aT\ll1$ regime the result
(\ref{59a}) for the entropy  holds. [Equation (\ref{59c}) includes the TM
mode as well.]

Now, $\varepsilon$ usually depends on $\zeta$. But this will not change
our conclusions from Eq.~(\ref{59a}).
To see this we can go back to expression (\ref{7}) which followed
from expansion of the logarithmic term in the free energy (\ref{5a}).
In the general case, the coefficients $A_m$ and $B_m$, which are less than 1,
should be included in Eq.~(\ref{7}), and 
powers of them will occur in the evaluation of the terms
contributing to the free energy for $k>1$.
These factors will all be smooth functions of $\zeta$ except for the case
of an idealized metal where $B_m$ becomes discontinuous at $\zeta=0$.
This smoothness is also valid for the Drude formula discussed in Appendix
\ref{VIa}.
With $A_m$ and $B_m$ included, Eq.~(\ref{5a}) can be summed with respect
to $\zeta_m=2\pi m/\beta$, and the Euler-Maclaurin formula (\ref{22})
can again be applied. 
(Equation (\ref{7}) with $B_m$ included is not applicable in this situation as 
we remarked there because $\zeta\to0$ is of relevance here.)
If $\varepsilon$ stays finite when $\zeta\to0$ the result clearly will be
the same as that given above.  However, for a real metal where
$\varepsilon\to\infty$ as $\zeta\to0$ the situation is more subtle.  For the
case of an ideal metal considered in Sec.~\ref{II}, the first derivative
$f'(0)$ was zero while $f'''(0)$ of Eq.~(\ref{stefan}) was nonzero.
By similar application of the Euler-Maclaurin formula to the free energy
(\ref{5a}) instead of the force (\ref{3}), the same will be true.
For a real metal obeying the Drude dispersion relation (\ref{1})
(with $\nu\ne0$) the first derivative $f'(0)$ continues to be zero
due to the $\zeta$ dependence of $B_m$, $B_m\sim\zeta_m^2$,
according to Eq.~(\ref{64}).  Thus, quite generally, we expect a
$T^3$ (or $T^4$) correction to the free energy at sufficiently low
temperature.

\subsection{Gold as a numerical example}
\label{V.3}
Let us go back to Eq.~(\ref{3}) for the surface force density, making use of 
the best available experimental results for $\varepsilon(i\zeta)$ as 
input when calculating the coefficients $A_m$ and $B_m$. We choose gold 
as an example. Useful information about the real and imaginary parts, 
$n'$ and $n''$, of the complex permittivity $n=n'+in''$, versus the 
real frequency $\omega$,  is given in Palik's book \cite{palik98} 
and similar sources. The range of photon energies given in Ref.~\cite{palik98} 
is from 0.1 eV to $10^4$\, eV. (The conversion factor 
\begin{equation}
1 \,\textrm{eV}=1.519\times 10^{15}\; \textrm{rad/s} 
\label{87}
\end{equation}
is useful to have in mind.) When $n'$ and $n''$ are known  the permittivity 
$\varepsilon(i\zeta)$ along the positive imaginary frequency axis, which is 
a real quantity, can be calculated by means of the Kramers-Kronig relations.
\begin{figure}
\centering
\includegraphics[height=10cm]{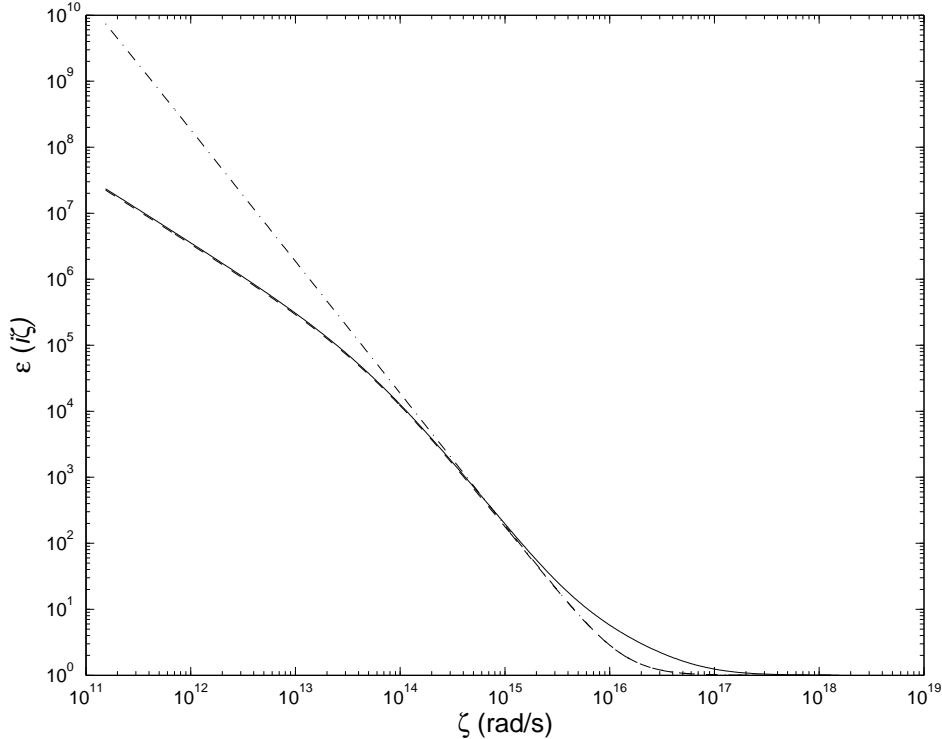}
\caption{Full line: Permittivity $\varepsilon (i\zeta)$ as function of imaginary 
frequency $\zeta$ for gold. The curve is calculated on the basis of 
experimental data.  Courtesy of Astrid 
Lambrecht and Serge Reynaud. Broken lines: $\varepsilon(i\zeta)$ versus 
$\zeta$ with $T$ as parameter, based upon the temperature dependent Drude
model; cf.~Appendix \ref{V.4}.  The upper curve is for $T=10$ K;
the lower is for $T=300$ K, which for energies below 1 eV ($1.5 \times 10^{15}$ rad/s)
nicely fits the experimental data.  Both curves are below the
experimental one for $\zeta>2\times 10^{15}$ rad/s.}
\label{fig1}
\end{figure}

Figure \ref{fig1} 
shows how $\varepsilon(i\zeta)$ varies with $\zeta$ over seven 
decades, $\zeta \in [10^{11}, 10^{18}]$ rad/s. The curve was given in an 
earlier paper \cite{lambrecht00}, and is reproduced here for convenience. 
(We are grateful to A. Lambrecht and S. Reynaud for having given us the 
results of their accurate calculations.)  At low photon
energies, below about 1 eV, 
the data are well described by the Drude model, Eq.~(\ref{1}), in which the 
input  parameters have the values \cite{lambrecht00}
\begin{equation}
\omega_p=9.0 \; \textrm{eV}, \quad \nu=35\; \textrm{ meV}.
\label{88}
\end{equation}
These values refer to room temperature. The curve in Fig.~\ref{fig1} shows a 
monotonic decrease of $\varepsilon(i\zeta)$ 
with increasing  $\zeta$, as any permittivity along the positive imaginary axis
has to follow according to thermodynamical requirements. The two broken 
curves in the figure show, for comparison, how $\varepsilon (i\zeta, T)$ varies
 with frequency if we accept the Drude model for all frequencies, and include 
 the temperature dependence of the relaxation frequency with $T$ as a 
 parameter.  Cf.~Appendix~\ref{V.4}. For $T=300$ K, the Drude curve is seen to be good 
 for all frequencies up to $\zeta \sim 2\times 10^{15}$ rad/s; for higher 
 $\zeta$ it gives too low values of $\varepsilon$. Both Drude curves, for 
 $T=10$ K and $T=300$ K, are seen to give the same values when $\zeta \ge 
 3\times 10^{14}$ rad/s.

The structure of Eq.~(\ref{3}) shows that for numerical integration it is 
advantageous to introduce the nondimensional quantity
\begin{equation}
y=qa
\end{equation}
as the integration variable. The force expression then takes the form
\begin{equation}
{\mathcal{F}}^T= -\frac{1}{\pi \beta a^3}{\sum_{m=0}^\infty}{}' 
\int_{m\gamma}^\infty y^2dy
\left[ \frac{A_me^{-2y}}{1-A_me^{-2y}}+\frac{B_me^{-2y}}{1-B_me^{-2y}} \right].
\label{90} 
\end{equation}
(This formula holds even  when practical units are restored, 
 when $\beta=1/k_BT$.) 
Typical magnitudes of the attractive pressure are about one millipascal, 
for a gap width of 1 $\mu$m. (The force between ideal metal plates at zero
temperature for 1 $\mu$m separation is 1.30 mPa.)

The next task is to determine the values of $A_m$ and $B_m$, in the limiting 
case of $m \rightarrow 0$. This has to be done analytically. Whereas the TM 
mode leads unambiguously to $A_0=1$ ($\varepsilon\gg1$), 
the TE mode is more delicate.  In Sec.~\ref{VIa}
 we show explicitly, by means of a limiting procedure based on the Drude 
 model, how $B_m \rightarrow 0$ when $\zeta \rightarrow 0$, i.e., when 
 $m \rightarrow 0$.  The $m=0$ TE mode 
 accordingly does not contribute. To summarize:
 \begin{subequations}
\begin{eqnarray}
 A_0&=&1,\quad B_0 =0\quad \textrm{for a metal}\quad(\varepsilon(0)=\infty),
 \label{91a}\\
A_0&=&\left( \frac{\varepsilon -1}{\varepsilon+1}\right)^2,\quad B_0=0 \quad 
\textrm{for a dielectric medium}\quad (\varepsilon=\varepsilon(0)).
\label{91}
\end{eqnarray}
\end{subequations}
These relations will be assumed in the following.

There are some general properties of the expression (\ref{90}) that ought to be 
noticed. First, at the lower limit, $y=m\gamma $, the coefficients $A_m$ 
and $B_m$ for $m\geq 1$ become equal,
\begin{equation}
A_m=B_m=\left( \frac{\sqrt{\varepsilon}-1}{\sqrt{\varepsilon}+1}\right)^2,
\quad\varepsilon=\varepsilon(i\zeta_m).
\label{92}
\end{equation}
This expression is precisely the reflection coefficient for Poynting's vector, 
at normal incidence. This special case obviously corresponds to $\mathbf{  k}_\perp 
=\mathbf{  0}$. Then the TE and TM modes are identical to each other. Secondly, we note 
that for large values of $y$, the integrand in Eq.~(\ref{90}) approaches
\begin{equation}
\left( \frac{\varepsilon -1}{\varepsilon +1}\frac{y}{e^y}\right)^2,
\quad\varepsilon=\varepsilon(i\zeta_m),
\label{93}
\end{equation}
showing how quickly the contributions from large $y$ die out.

\begin{figure}
\centering
\includegraphics[height=10cm]{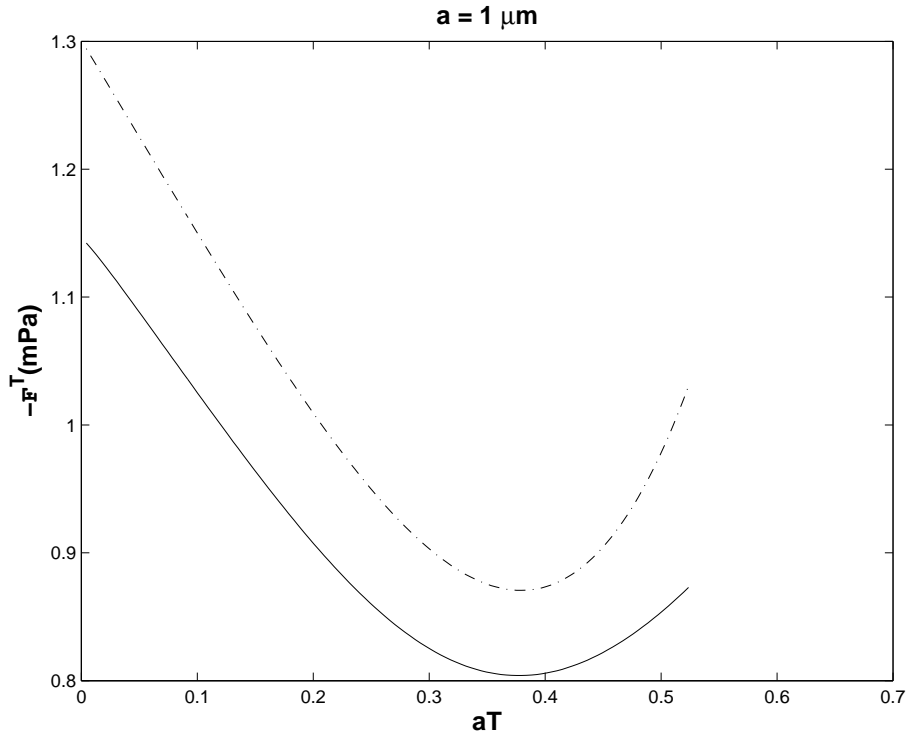}
\caption{Magnitude of surface force density for gold, in the temperature 
interval $10\,K \leq T \leq 1200$ K, when $a=1\; \mu$m. Solid line is physical 
result calculated from Eq.~(\ref{90}) where the room-temperature data for 
$\varepsilon(i\zeta)$ shown in Fig.~\ref{fig1} are used. Broken line is calculated 
from the ideal low-temperature form (\ref{linterm}).}
\label{fig2}
\end{figure}

The full line in Fig.~\ref{fig2} shows how the magnitude of ${\mathcal{F}}^T$ 
for gold 
varies with the dimensionless parameter $aT$, when $a=1\,\mu$m. The lower 
limit $aT=4.4\times 10^{-3}$ corresponds to the low temperature of $T=10$ K. 
Terminating the $y$ integration at the upper limit $y_\textrm{max}=30$ we found
the  necessary number of terms in the $m$ sum to be about $N=450$. At room 
temperature, $T=300$ K, corresponding to  $aT=0.131$ and $\gamma=0.823$, the 
required number of terms was found to be lower, $N=15$ (assuming the same 
$y_\textrm{max}$). In the upper limit, $aT=0.52$ ($T=1200$ K), only $N=4$ was 
required. This property of only a small number of terms being necessary at 
high temperatures is as we would expect. Note, however, that the temperature 
variation of $\varepsilon(i\zeta)$ is not taken into account. The only known 
empirical data for $\varepsilon(i\zeta)$ are referring to room temperature, 
and are as given in Fig.~\ref{fig1}.

The broken line in the same figure gives the result calculated from the
expression in Eq.~(\ref{linterm}),  which is for the modified ideal metal 
model in which the TE zero mode has been removed. 
The deviations from the full lines are seen to be 
quite uniform: 13\% at the lower limit, 12\% at room temperature, and 18\% 
at the upper limit. This uniformity in the deviations is somewhat surprising, 
 in view of the fact 
 that the expression (\ref{linterm}) is a low-temperature expansion which one 
would expect to be most accurate when $aT \rightarrow 0$. The reason for the 
deviations must lie in the different ways the two force expressions are 
calculated: Eq.~(\ref{linterm}) is based upon the idealized assumptions 
$A_m=B_m=1$ for 
all $m$ except that $B_0=0$, whereas Eq.~(\ref{90}) is calculated using the 
realistic dispersive data from Fig.~1, plus Eq.~(\ref{91a}) in the case $m=0$.

\begin{figure}
\centering
\includegraphics[height=10cm]{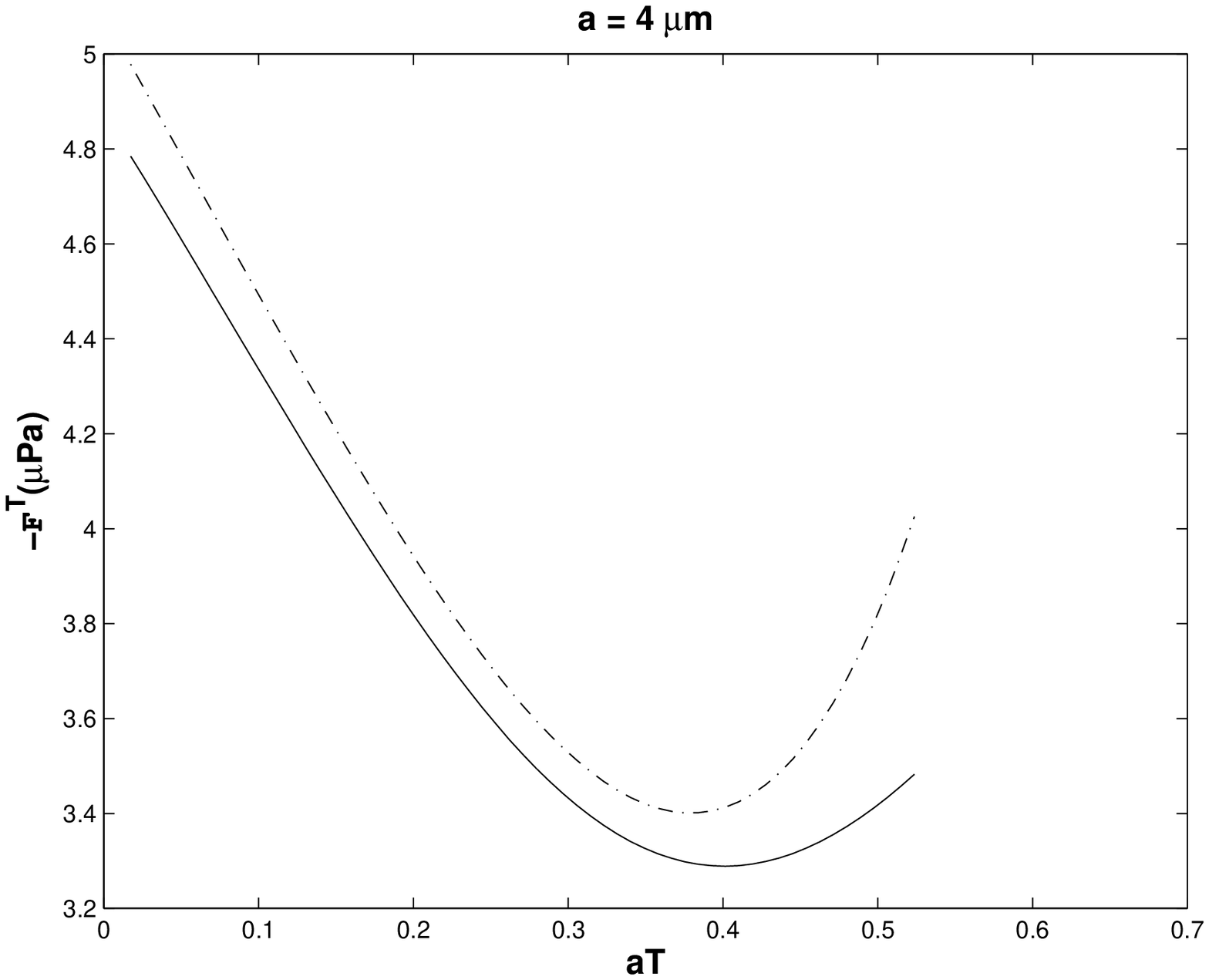}
\caption{Same as Fig.~\ref{fig2}, but at a larger spacing, $a=4\; \mu$m, 
corresponding to $10\,K\leq T\leq 300\,$ K.}
\label{fig3}
\end{figure}

Figure \ref{fig3}
 shows that the behaviour is essentially the same if the gap is made 
wider, $a=4\,\mu$m. The forces are now only about 0.4\% of those in 
Fig.~\ref{fig2}. 
The lower limit $aT=0.017$ corresponds to $T=10$ K ($N=115$ terms necessary), 
and the upper limit $aT=0.523$ corresponds to $T=300$ K ($\gamma=3.29$, $N=4$).
The deviations between the full dispersive result and Eq.~(\ref{linterm})
are now smaller than previously, about 5\%.

\begin{figure}
\centering
\includegraphics[height=10cm]{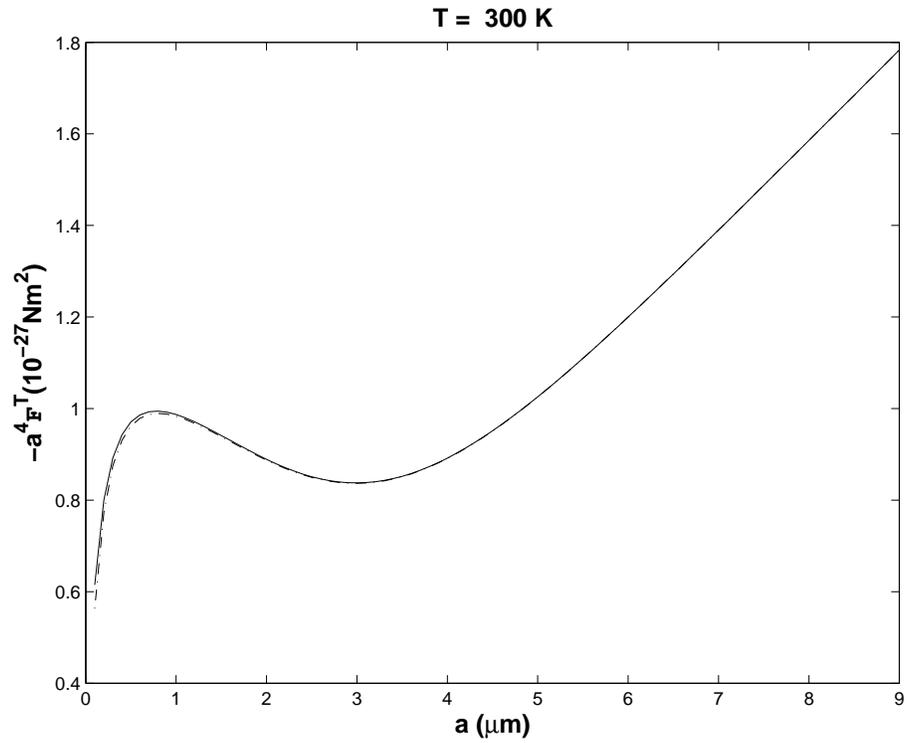}
\caption{ Surface force density for gold, multiplied with $a^4$, versus $a$ 
when $T=300$ K. Input data for $\varepsilon(i\zeta)$ are taken from 
Fig.~\ref{fig1}.}
\label{fig4}
\end{figure}

As experiments are usually made at room temperature for various gap widths, we 
show in Fig.~\ref{fig4} how the surface force density for gold varies with $a$,  
at $T=300$ K. We have here chosen to multiply the ordinate with $a^4$.
 The linear slope seen for $a\ge 4 \mu$m is nearly that predicted in
Eq.~(\ref{59c}), which gives a slope of $2.0\times 10^{-28}$ Nm$^2$/$\mu$m.
The linear region between 1 and 2 $\mu$m corresponds to that in 
Eq.~(\ref{linterm}) or (\ref{59c}) (intermediate temperatures).
Also shown is the prediction of the temperature dependent Drude model 
(Appendix \ref{V.4}),
when $T=300$ K. The differences are seen to be very small.  Since the Drude 
values for the permittivity are lower than the empirical ones at high 
frequencies, as seen in 
Fig.~\ref{fig1}, we expect the predicted Drude forces to be slightly 
weaker than those based upon the empirical permittivities. This expectation is 
borne out in Fig.~\ref{fig4}; the differences being large enough to be slightly visible 
at short distances, as we would expect since the plasma nature of the material 
becomes more pronounced for small distances.
Note that the temperature dependence of the permittivity is irrelevant here
because the temperature is fixed, unlike in Figs.~\ref{fig2} and \ref{fig3}.

\begin{figure}
\centering
\includegraphics[height=10cm]{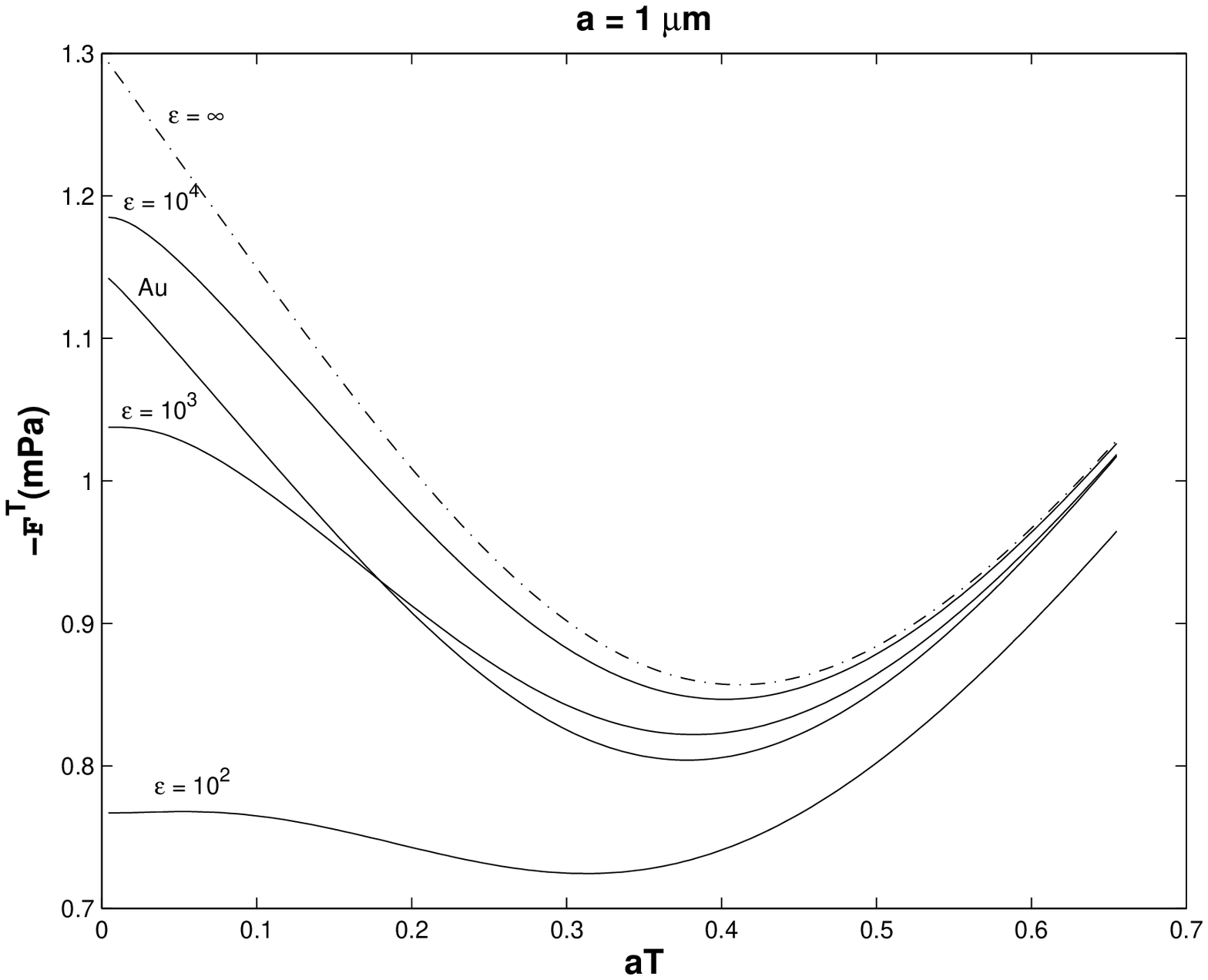}
\caption{Nondispersive theory: Surface force density calculated from 
Eq.~(\ref{90})
 for $\varepsilon \in\{100, 1000, 10000, \infty \}$.  The 
 $\varepsilon = \infty $ result is calculated from Eq.~(\ref{94}). 
 For low values of $aT$ the latter coincides with the expression (\ref{linterm}) 
 used in Fig.~\ref{fig2}. Also shown for comparison is the dispersive 
 result for gold, where experimental
  input data for $\varepsilon(i\zeta)$ are taken from 
 Fig.~\ref{fig1}. Gap width is $a=1\, \mu$m. The constraint $a=1\,\mu$m 
 applies only to the dispersive case, since otherwise $a^4{\mathcal F}^T$
 is a function of $aT$ only.}
 \label{fig5}
 \end{figure}

It is of interest to check the magnitude of the dispersive effect in these 
cases. We have therefore made a separate calculation of the expression 
(\ref{90}) 
when $\varepsilon$ is taken to be  constant. Figure \ref{fig5} 
shows how the force 
varies with $aT$ in cases when $\varepsilon\in\{100, 1000, 10000, \infty \}$ are 
inserted in the expressions for $A_m$ and $B_m$ in Eq.~(\ref{4a}).  Note that the 
$\varepsilon = \infty$ curve is obtained easily via the analytic result (\ref{16}), 
with $A_m=B_m=1$ for all $m\ge 1$. With $B_0=0$, Eq.~(\ref{16}) is modified into
\begin{equation}
{\mathcal   F}^T(\varepsilon =\infty)=\frac{1}{8\pi \beta a^3}
\left\{ \zeta(3)-\sum_{k=1}^\infty
\frac{1}{k^3}\, \left[ s_2(\gamma k)+2s_1(\gamma k)+2s_0(\gamma k)\right] 
\right\},
\label{94}
\end{equation}
which amounts to adding the last term of Eq.~(\ref{linterm}).
[The sum is alternatively given in Eq.~(\ref{42}), and the low-temperature
limit is given in Eq.~(\ref{linterm}).]
It is seen from the figure that the three first curves asymptotically 
approach the $\varepsilon =\infty$ curve, given by Eq.~(\ref{94}) when 
$\varepsilon$ increases, as we would expect. Again, we emphasize that the 
dispersive curve for gold is calculated using the available room-temperature 
data for $\varepsilon(i\zeta)$ from Fig.~\ref{fig1}. In the nondispersive case, 
there is of course no permittivity temperature problem since $\varepsilon$ is 
taken to be the same for all $T$.

There are several points worth noticing from Fig.~\ref{fig5}: 
(i) The curves have 
a horizontal slope at $T=0$. For finite $\varepsilon$ this property is 
clearly visible on the curves. This has to be so on physical grounds: 
If the force had a linear dependence on $T$
for small $T$ so would the free energy $F$, in contradiction with the
requirement that 
the entropy $S=-\partial F/\partial T$ has to go to zero as 
$T \rightarrow 0$. For the gold data the initial horizontal slope is
not resolvable on the scale of this graph, see the discussion at the end
of Sec.~\ref{V.2}.
 (ii) The curves show that the magnitude of the
force \textit{diminishes} 
with increasing $T$ (for a fixed $a$), in a certain temperature interval up 
to $aT \simeq 0.3$. This perhaps counterintuitive effect is thus clear from 
the nondispersive curves as well as from the dispersive curves in
 Figs.~\ref{fig2} and \ref{fig3}. (iii) It is seen that the curve for 
 $\varepsilon = \mbox{const.} = 1000$ gives a reasonably good approximation to 
 the real dispersive curve for gold when $a=1~\mu$m; the deviations are less 
  than about 5\% except for the lowest values of $aT$ ($aT <0.1$). 
This fact makes our neglect of the temperature dependence of 
$\varepsilon(i\zeta)$ appear physically reasonable; the various curves turn 
out to be rather insensitive with respect to variations in the input values of 
$\varepsilon(i\zeta)$.
  (iv)  One notes that the curves (for large $\varepsilon$) in Fig.~\ref{fig5} 
  are consistent with the free energy (\ref{59c}) using the rough 
  approximation (\ref{54}) for $B_m$. Especially one notes the initial 
  decrease of the magnitude of
  the Casimir force for increasing $T$ when $\varepsilon$ is 
  large. As discussed below Eq.~(\ref{59a}) this is again connected with the 
  counterintuitive negative contribution to the entropy.  
  (v)  Also, it can be remarked that $B_0=0$ is required when 
  $\varepsilon$ is finite. Otherwise the curves in Fig.~\ref{fig5}, and thus 
  the free energy, would have a finite slope at $T=0$ which again would 
  imply a finite entropy contribution at $T=0$ in violation with the third 
 law of thermodynamics.

\begin{table}
\centering
\begin{tabular}{|ccccccc|} \hline\hline
\multicolumn{3}{c}{}           & \multicolumn{2}{c}{$y=1$} & \multicolumn{2}{c}{$y=3$} \\ \hline
$m$& $\varepsilon(i\zeta_m) \times 10^3$  &$\zeta_m\times10^{12}$rad/s&$A_m$    &$B_m$  &$A_m$  & $B_m$ \\ \hline
1  &  382.0                                      &      8.226            &0.9998  &0.7899 &0.9999 & 0.4944 \\
3  &  100.4                                      &     24.68             &0.9990  &0.8578 &0.9997 & 0.6317 \\ 
5  &  49.76                                      &     41.13             &0.9975  &0.8774 &0.9992 & 0.6759 \\
7  &  30.28                                      &     57.58             &0.9956  &0.8872 &0.9985 & 0.6985 \\
9  &  20.52                                      &     74.03             &0.9931  &0.8930 &0.9977 & 0.7124 \\
11 &  14.87                                      &     90.49             &0.9902  &0.8970 &0.9967 & 0.7219  \\
13 &  11.30                                      &     106.9             &0.9867  &0.8998 &0.9955 & 0.7288 \\ 
15 &  8.891                                      &     123.4             &0.9827  &0.9020 &0.9942 & 0.7341 
\\ \hline  \hline 
\end{tabular}
\caption{Some data in the dispersive theory for gold.  Here $T=10$ K,
$y\equiv qa\in\{1,3\}$. Room temperature input data for $\varepsilon (i\zeta)$ 
are taken from Fig.~\ref{fig1}.} 
\label{tabI}
\end{table}

\begin{table}\centering
\begin{tabular}{|ccccccccc|} \hline\hline
$a(\mu\textrm{m})$  & $m=0$  & $m=1$  & $m=2$  &  $m=3$  & $m=4$  & $m=5$  & $m=6$   & $m=7$    \\ \hline 
0.5              & 0.32   &  0.98  & 1.03   &  1.05   & 1.06   & 1.07   & 1.08    & 1.08  \\  
1                & 0.58   & 1.98   & 2.05   & 2.07    & 2.08   & 2.08   & 2.07    & 2.06  \\
2                & 1.10   & 4.04   & 4.09   & 4.07    &4.02    & 3.96   & 3.88    & 3.79 \\
3                & 1.63   & 6.11   & 6.09   & 5.98    & 5.80   & 5.59   & 5.36    & 5.10 \\
4                & 2.16   & 8.18   & 8.04   & 7.75    & 7.37   & 6.93   & 6.45    & 5.95 \\
5                & 2.69   & 10.24  & 9.92   & 9.37    & 8.69   & 7.94   & 7.16    & 6.38 \\
6                & 3.23   & 12.30  & 11.71  & 10.81   & 9.75   & 8.63   & 7.51    & 6.45 \\
7                & 3.78   & 14.33  & 13.39  & 12.06   & 10.55  & 9.02   & 7.56    & 6.24
\\ \hline \hline
\end{tabular}  
\caption{Contribution from the various Matsubara frequencies for gold.
What is given is the percentage of ${\mathcal   F}^T$ for each mode in the region
$m\in [0,7]$.  The temperature is $T=10$ K. Room temperature input data for 
$\varepsilon(i\zeta)$ are taken from Fig.~\ref{fig1}.}
\label{tabII}
\end{table}
\begin{table}
\centering
\begin{tabular}{|ccccccccc|} \hline\hline
$a(\mu\textrm{m})$  & $m=0$  & $m=1$  & $m=2$  &  $m=3$  & $m=4$  & $m=5$  & $m=6$   & $m=7$    \\ \hline 
0.5              & 10.20  & 31.24  & 22.95  &  15.09  &  9.18  & 5.28   & 2.91    & 1.55 \\ 
1                & 20.07  & 49.37  & 20.83  & 6.97    & 2.03   & 0.54   & 0.14    & 0.03 \\ 
2                & 44.56  & 49.87  & 5.17   & 0.37    & 0.02  &&&\\
3                & 70.95  & 28.41  & 0.63   & 0.01 &&&&\\
4                & 88.88  & 11.07  & 0.05  &&&&&\\
5                & 96.58  & 3.42 &&&&&&\\
6                & 99.06  & 0.94 &&&&&&\\
7                & 99.76  & 0.24&&&&&&
\\ \hline\hline
\end{tabular}

\caption{Same as in Table \ref{tabII}, but at temperature 300 K. Data
from Fig.~\ref{fig1} is again used.}
\label{tabIII}
\end{table}

Instead of confining ourselves to a ``black box" calculation of 
the force expression (\ref{90}), it is desirable to break up the 
expression somewhat, to see how the various values of $m$ contribute. 
We do this in Tables \ref{tabI}--\ref{tabIII}, for gold. 
The first two tables refer to the case $T=10$ K.
(Again, the experimental values of $\varepsilon(i\zeta)$
at room temperature are used.) As $y$ is the important 
integration parameter in Eq.~(\ref{90}), we keep $y$ fixed in Table \ref{tabI},
 $y\in\{1,3\}.$  It is seen that $A_m$ stays close to 1, whereas $B_m$ decreases 
 for increasing $y$, if $m$ is kept constant. Table \ref{tabII} shows how the 
 various $m$ contribute to the force. Writing the total force as a sum,
\begin{equation}
 {\mathcal   F}^T=\sum_{m=0}^\infty {\mathcal   F}_m^T ,
\end{equation}
the columns in the table show the percentage of ${\mathcal   F}_m^T$, i.e., 
$({\mathcal   F}_m^T/{\mathcal   F}^T)\times 100$, distributed over the region $m 
\in [0,7]$, when $T=10$ K. The distribution from the various $m$s is 
seen to be very broad, as is characteristic for a low-temperature problem. 
Table \ref{tabIII} shows the same kind of distribution over $m$ when 
$T=300$ K. Already from a gap distance of $a$ = 3--4 $\mu$m onwards, 
the distribution is heavily concentrated around low $m$, as is characteristic 
of a high-temperature problem. 

It is in this context instructive as a corollary to go back to the integral 
over $y$ in Eq.~(\ref{90}). One would expext the main contribution to the 
integral to come from the region $y=qa=\sqrt{k_\perp^2+\zeta^2}\, a \sim 1$. 
Assuming the most important values of $ k_\perp $ to be moderate, this means 
$\zeta a \sim 1$, or $m \sim 1/(2\pi aT)$, since $\zeta =2\pi mT$. When $T=300$
 K, we thus expect the dominant contribution to come from $m\sim 1$ when $a=1\,
\mu$m, and from $m=0$ when
$a \geq 3\,\mu$m. This is seen to agree very well with the data in Table 
\ref{tabIII}. Similar considerations apply to the case $T=10$ K, although 
the contributions from the various $m$s are then more smeared out.

The important question is now: Have the characteristic temperature variations 
shown in the theoretical figures above been verified in practice? Of most 
interest in this context is the experiment of Bressi et al.~\cite{bressi02}, 
since it deals with parallel plates directly. According to personal 
information from R. Onofrio, one of the members of the Italian group, 
the observed Casimir forces were lower than those predicted by the traditional
(SDM) theory for conducting plates, in cases where  the distances were low, $a 
\leq 0.5 \, \mu $m. This reduction effect is apparent also from their Fig.~4. 
Now, the plates in this experiment were coated with chromium rather than with 
gold, but we can check that the corrections in that case
are of the same magnitude as if the 
plates were coated with gold. Namely, an explicit calculation of the analogue of 
Fig.~\ref{fig5} for the case $a=0.5\,\mu$m (not shown here) shows that at 
room temperature for which $aT=0.065$, the force becomes $-{\mathcal   F}^T= 
15.5$ mPa. The conventional (SDM) theory gives in this case the force 
$1.3\times 2^4=20.8$ mPa. The predicted reduction in the force is thus about 
25\%, somewhat more than the measurements indicate. In any case, this 
suggests that the reduced force seen at room temperature in Ref.~\cite{bressi02}
may be the first actual observation of the  temperature effect predicted 
theoretically.

At larger distances, however, between 1 and 2 $\mu$m, the situation is no 
longer so clear-cut, since they observe a Casimir force in excess of the 
theoretically predicted one. The reason for this deviation is not known. Of 
course the force becomes weaker at larger distances, thus being subject to 
larger experimental uncertainties. The most natural conclusion to be drawn at this 
stage is that we have to wait for better precision in this kind of 
difficult experiment.  Ideas for such an improved experiment which could
descriminate between the different models have just appeared \cite{ckmm}.

\appendix

\section{On the smoothness of the reflection coefficient $r_2$ at
low frequencies, for a metal}
\label{VIa}
In view of the current discussion in the literature about the value of
the reflection coefficient $r_2$ for a metal in the limit of low
frequencies, let us consider this point in some detail. As mentioned
earlier, the problem occurs in connection with use of the Drude
formula, Eq.~(\ref{1}). The coefficient $r_2$ is actually the square root of
our quantity $B_m$ defined in Eq.~(\ref{4a}), so that we may write
\begin{equation}
r_2^2=\left( \frac{s-p}{s+p}\right)^2.
\label{60}
\end{equation}
Let us keep the transverse wave vector $\mathbf{  k}_\perp$ fixed, and
perform a power series expansion of $\varepsilon(i\zeta)$ to the first order
in $\zeta/\nu$. [Any normal metal must have a finite relaxation
frequency $\nu$, so that in the limit of low frequencies, $\zeta/\nu$
can be regarded as small.
At zero temperature, we are assuming $\nu(T=0)\ne0$.] From Eq.~(\ref{1}) we get
\begin{equation}
\varepsilon(i\zeta)-1 \rightarrow \frac{\omega_p^2}{\nu \zeta}\left(
1-\frac{\zeta}{\nu}\right),
\label{61}
\end{equation}
which for the Lifshitz variables $s$ and $p$ implies [cf. Eq.~(\ref{4b})]
\begin{subequations}
\begin{eqnarray}
 s&=&\sqrt{\varepsilon-1+p^2}\rightarrow \frac{k_\perp}{\zeta}\left(
1+\frac{\omega_p^2\zeta}{2\nu k_\perp^2}\right), 
\label{62}
\\
p&=&\frac{k_\perp}{\zeta}\sqrt{1+\frac{\zeta^2}{k_\perp^2}}
\rightarrow \frac{k_\perp}{\zeta}.\label{63}
\end{eqnarray}
\end{subequations}
Insertion into Eq.~(\ref{60}) now yields
\begin{equation}
r_2^2\rightarrow \left(\frac{\omega_p^2}{4k_\perp^2
}\right)^2\left(\frac{\zeta}{\nu}\right)^2.
\label{64}
\end{equation}
We thus see that $r_2^2 \rightarrow 0$ smoothly as $\zeta\rightarrow
0$.  Contrary to recent statements in the literature
\cite{klimchitskaya01,bordag00,fischbach01}, we find that there is no
peculiar effect taking place at $\zeta=0$, when the Drude model is
used. The result (\ref{64}) corresponds to a vanishing contribution to the
Casimir effect from the $m=0$ TE mode for a real metal, in accordance
with our treatment in Sec.~\ref{V}.

The argument above hinged on the assumption that $\mathbf{  k}_\perp \neq
\mathbf{  0}$. 
One might wonder: What happens if $\mathbf{  k}_\perp$ is exactly zero? 
Mathematically, it then follows from Eq.~(\ref{4b}) that $r_2^2=1$. 
This case cannot, however, be of physical importance. 
The set $\mathbf{  k}_\perp=\mathbf{  0}$ is mathematically of measure zero, 
and has thus no influence upon real physics.

\section{On the physical importance of $A_m$ and $B_m$ }
\label{VIb}

It is physically instructive to show in some detail how the coefficients 
$A_m$ and $B_m$ relate to the conventional Fresnel coefficients in optics, 
at oblique incidence. Consider first the TM mode, and let a plane wave be
 incident from the left (medium 1, refractive index $n_1=\sqrt{\varepsilon }$) 
 at a real angle of incidence $\theta_i$ towards the boundary located at $z=0$.
  The angle of transmission to the vacuum region $z>0$ is $\theta_t$. For 
  instance from Ref.~\cite{born80} we have for the  ratio between the 
  reflected wave amplitude $R^\textrm{TM}$ and the incident wave amplitude 
  $A^\textrm{TM}$
\begin{subequations}
\begin{equation}
\frac{R^\textrm{TM}}{A^\textrm{TM}}=\frac{\cos {\theta_i}-n_1 \cos{\theta_t}}
{\cos {\theta_i}+n_1\cos {\theta_t}}.
\label{99}
\end{equation}
Since $\cos {\theta_i}=\sqrt{1-k_\perp ^2/(\varepsilon \omega^2)}$,
 $\cos{\theta_t}=\sqrt{1-k_\perp^2/\omega^2}$ we get, when replacing 
 $\omega$ by $i \zeta$,
\begin{equation}
\frac{R^\textrm{TM}}{A^\textrm{TM}}=
\frac{\sqrt{\varepsilon +k_\perp^2/\zeta^2}-\varepsilon 
\sqrt{1+k_\perp^2/\zeta^2}}
{\sqrt{\varepsilon+k_\perp^2/\zeta^2}+\varepsilon \sqrt{1+k_\perp^2/\zeta^2}}.
\label{100}
\end{equation}
\end{subequations}
Now $s=\sqrt{\varepsilon -1+p^2}=\sqrt{\varepsilon +k_\perp^2/\zeta^2}$,  
$p=q/\zeta=\sqrt{1+k_\perp^2/\zeta^2}$, and so we get
\begin{subequations}
\begin{equation}
\frac{R^\textrm{TM}}{A^\textrm{TM}}=\frac{s-\varepsilon p}{s+\varepsilon p}
=\sqrt{A_m}.
\label{101}
\end{equation}
Similarly for the TE mode,
\begin{equation}
\frac{R^\textrm{TE}}{A^\textrm{TE}}=\frac{s-p}{s+p}=\sqrt{B_m}.
\label{102}
\end{equation}
\end{subequations}
Of course, these results are also found in textbooks \cite{ce}.

\section{Parallel dielectrics}
\label{VIc}
In Ref.~\cite{milton01} the following result for the TE reduced Green's 
function is given,
\begin{equation}
g^H(z,z')
=\frac1{2\kappa_2}\left(e^{-\kappa_2|z-z'|}+r\,e^{-\kappa_2(z+z'-2a)}\right).
\end{equation}
which is valid for $z$, $z'>a$.  Here the reflection coefficient is
\begin{subequations}
\begin{eqnarray}
r&=&\frac{\kappa_2-\kappa_3}{\kappa_2+\kappa_3}+\frac{4\kappa_2\kappa_3}
{\kappa_3^2-\kappa_2^2}d^{-1},\\
d&=&\frac{\kappa_3+\kappa_1}{\kappa_3-\kappa_1}
\frac{\kappa_3+\kappa_2}{\kappa_3-\kappa_2}e^{2\kappa_3a}-1,
\end{eqnarray}
\end{subequations}
 and 
\begin{equation}
\kappa_i^2=k^2-\omega^2\epsilon_i,
\end{equation}
and we have taken a parallel dielectric slab geometry
\begin{equation}
\epsilon(z)=\left\{\begin{array}{cc}
\epsilon_1,&z<0,\\
\epsilon_3,&0<z<a,\\
\epsilon_2,&a<z.
\end{array}\right.
\end{equation}
The temperature controversy centers on the zero mode.  If $\omega^2\epsilon$
vanishes at $\omega=0$ (true for the
Drude model, but not the plasma model), 
then the reflection coefficient vanishes there,
$r=0$, and we have only a free Green's function at 
$\omega=0$, that is, the boundary
becomes transparent.  The TM reflection coefficient does not have this
property.

We have redone the calculation to find the reduced Green's function in the
interior region, $0<z,z'<a$.  We find
\begin{eqnarray}
g^H(z,z')&=&\frac1{2\kappa_3}\bigg\{e^{-\kappa_3|z-z'|}+\frac{\kappa_3-\kappa_1}
{\kappa_3+\kappa_1}e^{-\kappa_3(z+z')}
+d^{-1}\bigg[e^{\kappa_3(z-z')}+e^{\kappa_3(z'-z)}\nonumber\\
&&\qquad\mbox{}+\frac{\kappa_3+\kappa_1}{\kappa_3-\kappa_1}e^{\kappa_3(z+z')}
+\frac{\kappa_3-\kappa_1}{\kappa_3+\kappa_1}
e^{-\kappa_3(z+z')}\bigg]\bigg\}.
\label{gh}
\end{eqnarray}
Again, it is easy to see that we  obtain only the free Green's function for
the zero mode:
\begin{equation}
g^H(z,z';\omega=0)=\frac1{2k}e^{-k|z-z'|},
\end{equation}  
provided $\lim_{\omega\to0}\omega^2\epsilon(\omega)=0$.

A check of this result is that if we substitute Eq.~(\ref{gh}) into the expression
for the force/area (3.13) of Ref.~\cite{milton01}, 
we get for the TE contribution to the force [see Eq.~(3.10) there]
\begin{eqnarray}
{\mathcal   F}^T&=&\frac{i}{2}\int\frac{d\omega}{2\pi}\frac{(d\mathbf{  k})}{(2\pi)^3}
(\epsilon_2-\epsilon_3)\omega^2g^H(a,a)\nonumber\\
&=&\frac{i}{2}\int\frac{d\omega}{2\pi}\frac{(d\mathbf{  k})}{(2\pi)^3}
\left(\kappa_3-\kappa_2+2\kappa_3 d^{-1}\right),
\end{eqnarray}
identical to the first term in Eq.~(3.19) of Ref.~\cite{milton01},
and apart from a contact term is the same as the second term in Eq.~(\ref{3}).
All of this does not seem to support the claims of Klimchitskaya et 
al.~\cite{klimchitskaya01,klim,bordag00,fischbach01}
that there is something ill-defined about the $\omega=0$ limit.\\

\section{Temperature dependence of the relaxation frequency for gold}
\label{V.4}
To investigate the temperature dependence of the relaxation frequency 
$\nu(T)$ in the Drude relation
\begin{equation}
\varepsilon(i\zeta, T)=1+\frac{\omega_p^2}{\zeta [\zeta +\nu(T)]}
\end{equation}
for gold, it is convenient to make use of the Bloch-Gr{\"u}neisen formula for 
the temperature dependence of the electrical resistivity $\rho$ 
\cite{handbook67}:
\begin{equation}
\rho(T)=C\left( \frac{T}{\Theta}\right)^5\int_0^{\Theta/T}\frac{x^5e^x\,dx}
{(e^x-1)^2}.
\end{equation}
It is known that $\Theta=175$ K for gold. The constant $C$ can be determined 
from the knowledge that $\rho= 2.20\times 10^{-8}
\, \Omega $ m at temperature 295 K \cite{handbook72}. We obtain $C=5.32\times 
10^{-8}\,\Omega\,$m.

\begin{figure}
\centering
\includegraphics[height=10cm]{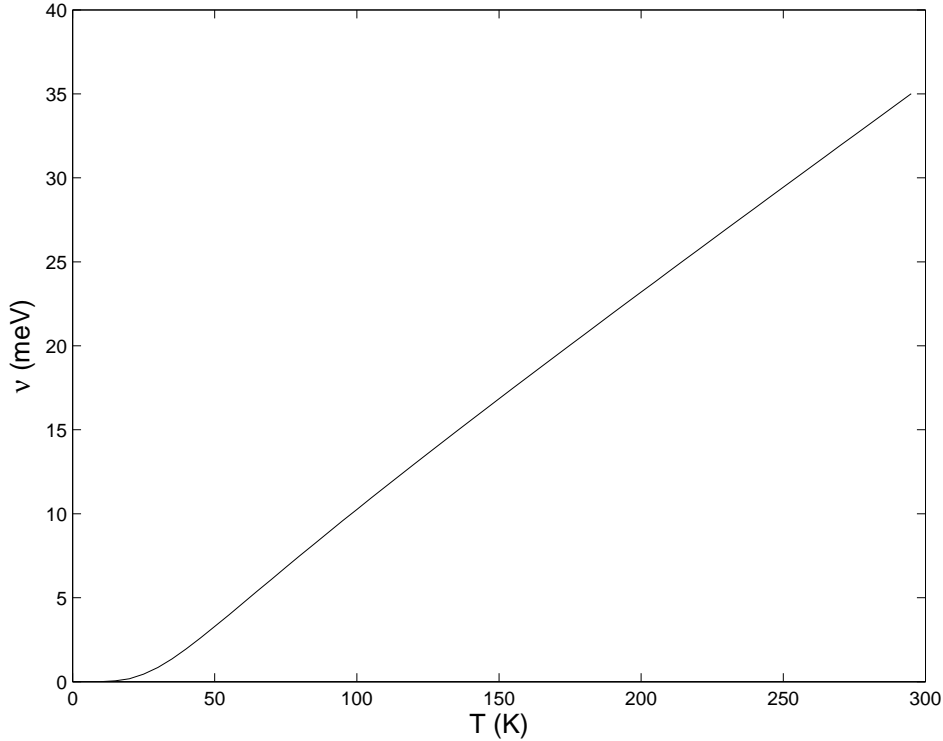}
\caption{Temperature dependence of the relaxation frequency for gold.}
\label{fig6}
\end{figure}

The theoretical relationship between $\nu$ and the static resistivity $\rho$ is
\begin{equation}
\nu= \frac{f_0N_e\,e^2}{m}\,\rho,
\end{equation}
where $N_e$ is the number density of atoms, $f_0 N_e$ with $f_0 \sim 1$ the 
number density of free electrons, and $m$ the effective electron mass. The 
simplest way to proceed is to put $\nu=K\rho $ with $K$ a constant, and make 
use of the room-temperature data of Eq.~(\ref{88}). We obtain $K=1.59\times 
10^6$ eV $\Omega^{-1}\,$m$^{-1}$. Altogether,
\begin{equation}
\nu(T)=0.0847\left(\frac{T}{\Theta}\right)^5\int_0^{\Theta/T}\frac{x^5e^x\,dx}
{(e^x-1)^2},
\end{equation}
where the unit of $\nu(T)$ is eV. The temperature variation is shown in 
Fig.~\ref{fig6}. For low temperatures, $\nu(T)\propto T^5$, whereas at high 
temperatures, $\nu(T) \propto T$. The curve is seen to be similar to the one 
given in Fig.~3 of Ref.~\cite{bezerra02}, in the case of aluminum.
 
An important caveat must be mentioned, however; these formulas neglect the
effect of impurities, which give rise to a nonzero resistivity at zero
temperature \cite{resist}.  This makes the use of these ideal 
resistivity models 
questionable, and adds further evidence that the behavior of the entropy
discussed in Sec.~\ref{V} is correct.  

\begin{acknowledgments}

KAM is grateful to the US Department of Energy for partial financial
support of this research.  He would like to thank Peter van Nieuwenhuizen
for discussion about the subtleties of zero modes. IB thanks Astrid Lambrecht 
and Serge Reynaud for providing their numerical calculations of the 
dispersion relation for gold, Roberto Onofrio for information about 
the experiment of Ref.~\cite{bressi02}, Vladimir 
Mostepanenko for valuable discussions on the relaxation frequency temperature 
problem, as well as for information about numerical data, and
Bo Sernelius for discussions about the resistivity of metals at very low
temperatures.
\end{acknowledgments}


\begin{thebibliography}{99}
\bibitem{milton01}
K. A. Milton, \textit{The Casimir Effect: Physical Manifestations of the
Zero-Point Energy} (World Scientific, Singapore, 2001).
\bibitem{mostepanenko97}
V. M. Mostepanenko and N. N. Trunov, \textit{The Casimir Effect and its
Applications} (Clarendon Press, Oxford, 1997).
\bibitem{milonni94}
P. W. Milonni, \textit{The Quantum Vacuum} (Academic Press, San Diego,
1994).
\bibitem{plunien86}
G. Plunien, B. M{\"u}ller, and W. Greiner, Phys. Reports \textbf{134}, 87
(1986).
\bibitem{bordag01}
M. Bordag, U. Mohideen, and V. M. Mostepanenko, Phys. Reports \textbf{353}, 
1 (2001).

\bibitem{klimchitskaya01}
G. L. Klimchitskaya and V. M. Mostepanenko, Phys. Rev. A \textbf{63},
062108 (2001).
\bibitem{bordag00}
M. Bordag , B. Geyer, G. L. Klimchitskaya, and V. M. Mostepanenko,
Phys. Rev. Lett. \textbf{85}, 503 (2000).
\bibitem{fischbach01}
E. Fischbach, D. E. Krause, V. M. Mostepanenko, and M. Novello, Phys.
Rev. D \textbf{64}, 075010 (2001).
\bibitem{bezerra02}
V. B. Bezerra, G. L. Klimchitskaya, and V. M. Mostepanenko, Phys. Rev. A
{\bf 66}, 062112 (2002).
\bibitem{lifshitz80}
E. M. Lifshitz and L. P. Pitaevskii, \textit{Statistical Physics, Part 2}
(Pergamon Press, Oxford, 1980), Sec.~81.
\bibitem{schwinger78}
J. Schwinger, L. L. DeRaad, Jr., and K. A. Milton, Ann. Phys. (N.Y.)
\textbf{115}, 1 (1978).
\bibitem{hoye01}
J. S. H{\o}ye, I. Brevik, and J. B. Aarseth, Phys. Rev. E \textbf{63},
051101 (2001).
\bibitem{bostrom00}
M. Bostr{\"o}m and Bo E. Sernelius, Phys. Rev. Lett. \textbf{84}, 4757
(2000).
\bibitem{brevik02}
I. Brevik, J. B. Aarseth, and J. S. H{\o}ye, Int. J. Mod. Phys. A 
\textbf{17}, 776 (2002).
\bibitem{brevik02a}
I. Brevik, J. B. Aarseth, and J. S. H{\o}ye, 
 Phys. Rev. E \textbf{66}, 026119 (2002).
\bibitem{lamoreaux00}
S. K. Lamoreaux, quant-ph/0007029 v4.
\bibitem{sernelius01}
Bo E. Sernelius and M. Bostr{\"o}m, Phys. Rev. Lett. \textbf{87}, 259101
(2001).
\bibitem{bordag01a}
M. Bordag, B. Geyer, G. L. Klimchitskaya, and V. M. Mostepanenko, Phys.
Rev. Lett. \textbf{87}, 259102 (2001).
\bibitem{lamoreaux97}
S. K. Lamoreaux, Phys. Rev. Lett. \textbf{78}, 5 (1997).
\bibitem{mohideen98}
U. Mohideen and A. Roy, Phys. Rev. Lett. \textbf{81}, 4549 (1998).
\bibitem{roy99}
A. Roy, C.-Y. Lin, and U. Mohideen, Phys. Rev. D \textbf{60}, R111101
(1999).
\bibitem{harris00}
B. W. Harris, F. Chen, and U. Mohideen, Phys. Rev. A \textbf{62}, 052109
(2000).
\bibitem{chen02}
F. Chen, U. Mohideen, G. L. Klimchitskaya, and V. M. Mostepanenko,
Phys. Rev. Lett. \textbf{88}, 101801 (2002).
\bibitem{ederth00}
T. Ederth, Phys. Rev. A \textbf{62}, 062104 (2000).
\bibitem{chan01}
H. B. Chan, V. A. Aksyuk, R. N. Kleiman, D. J. Bishop, and F. Capasso,
Phys. Rev. Lett. \textbf{87}, 211801 (2001); Science \textbf{291}, 1941
(2001).
\bibitem{bressi02}
G. Bressi, G. Carugno, R. Onofrio, and G. Ruoso, Phys. Rev. Lett. 
\textbf{88}, 041804 (2002).
\bibitem{lambrecht02}
A. Lambrecht and S. Reynaud, in \textit{Seminaire Poincare 1}
 (Institut Henri Poincare, Paris, 9 March 2002), pp.~79--92
[www.lpthe.jussieu.fr/poincare/], quant-ph/0302073.
\bibitem{blocki77}
J. Blocki, J. Randrup, W. J. Swialecki, and C. F. Tsang, Ann. Phys.
(N.Y.) \textbf{105}, 427 (1977).
\bibitem{lamoreaux98}
S. K. Lamoreaux, Phys. Rev. Lett. \textbf{81}, 5475 (1998); Phys. Rev. A
\textbf{59}, R3149 (1999).
\bibitem{lambrecht00}
A. Lambrecht and S. Reynaud, Eur. Phys. J. D \textbf{8}, 309 (2000); Phys.
Rev. Lett. \textbf{84}, 5672 (2000).
\bibitem{genet02}
C. Genet, A. Lambrecht, and S. Reynaud, Int. J. Mod. Phys. A \textbf{17},
761 (2002).
\bibitem{svetovoy00}
V. B. Svetovoy and M. V. Lokhanin, Mod. Phys. Lett. A \textbf{15}, 1437
(2000); quant-ph/0004004;  Phys. Lett. A \textbf{280}, 177 (2001).
\bibitem{barton01}
G. Barton, Phys. Rev. A \textbf{64}, 032103 (2001).
\bibitem{feinberg01}
J. Feinberg, A. Mann, and M. Revzen, Ann. Phys. (N.Y.) \textbf{288}, 103
(2001).
\bibitem{bezerra02a}
V. B. Bezerra, G. L. Klimchitskaya, and V. M. Mostepanenko,
Phys. Rev. A {\bf 65}, 052113 (2002).
\bibitem{torgerson} J. R. Torgerson and S. K. Lamoreaux, 
quant-ph/0208042.
\bibitem{ms} P. C. Martin and J. Schwinger, Phys.\ Rev.\ \textbf{115}, 1342
 (1959).
\bibitem{klim} G. L. Klimchitskaya, Int.\ J. Mod.\ Phys.\ A \textbf{17}, 751
(2002).

\bibitem{hoye98}
J. S. H{\o}ye and I. Brevik, Physica (Amsterdam) A \textbf{259}, 165
(1998).
\bibitem{sauer62}
F. Sauer, PhD thesis, G{\"o}ttingen, 1962.
\bibitem{mehra67}
J. Mehra, Physica (Amsterdam) \textbf{37}, 145 (1967).
\bibitem{hoye81}
J. S. H{\o}ye and G. Stell, J. Chem. Phys. \textbf{75}, 5133 (1981).
\bibitem{brevik88}
I. Brevik and J. S. H{\o}ye, Physica A \textbf{153}, 420 (1988).
\bibitem{brevik79}
I. Brevik, Phys. Reports \textbf{52}, 133 (1979).
\bibitem{palik98}
\textit{Handbook of Optical Constants of Solids}, edited by E. D. Palik 
(Academic Press, New York, 1998).

\bibitem{ckmm} F. Chen, G. L. Klimchitskaya, U. Mohideen, and
V. M. Mostepanenko, quant-ph/0302149, to appear in Phys. Rev. Lett.

\bibitem{born80}
M. Born and E. Wolf, \textit{Principles of Optics}, 6th ed. 
(Pergamon Press, Oxford, 1980), p. 40.

\bibitem{ce} J. Schwinger, L. L. DeRaad, Jr., K. A. Milton, and W.-y. Tsai,
\textit{Classical Electrodynamics\/} (Perseus, New York, 1998).

\bibitem{handbook67}
See, for instance,  \textit{Handbook of Physics}, edited by E. U. Condon 
and H. Odishaw, 2nd ed. (McGraw-Hill, New York, 1967), Eq.~(6.12).

\bibitem{handbook72}
\textit{American Institute of Physics Handbook}, edited by D. E. Gray, 
3rd ed. (McGraw-Hill, New York, 1972).
\bibitem{resist} M. Khoshenevisan, W. P. Pratt, Jr., P. A. Schroeder, and
S. D. Steenwyk, Phys. Rev. B {\bf 19}, 3873 (1979).
\end{thebibliography}
\end{document}